\documentclass[draft]{agujournal2019}
\usepackage{url} 
\usepackage[finalnew]{trackchanges} 
\usepackage{soul}

\usepackage{rotating} 
\usepackage{booktabs} 
\usepackage{array}    
\draftfalse

\journalname{Journal of Advances in Modeling Earth Systems (JAMES)}

\begin{document}
%
%


\title{The Equilibrium Response of Atmospheric Machine-Learning Models to Uniform Sea Surface Temperature Warming}

%
%




\authors{Bosong Zhang\affil{1}\thanks{300 Forrestal Rd, Princeton, NJ 08540} and Timothy M. Merlis\affil{1}}
\affiliation{1}{Program in Atmospheric and Oceanic Sciences, Princeton University, Princeton, NJ}




\correspondingauthor{Bosong Zhang}{bosongz@princeton.edu}



\begin{keypoints}

\item \remove{Machine learning models can stably simulate the present-day atmosphere over long timescales.} \add{Three distinct types of machine learning (ML) global atmosphere models are evaluated for their ability to simulate the present-day climate.}
\item \remove{Uniform sea surface temperature warming serves as a benchmark for evaluating ML models' climate change responses.} \add{The three ML models show mixed agreement with a physics-based atmosphere model across key climate metrics in response to uniform surface warming.}
\item \remove{Machine learning models show mixed agreement with a physics-based atmosphere model across key climate metrics.} \add{Fully data-driven models struggle to capture physically consistent climate change signals, indicating that further improvements are needed for robust out-of-sample generalization.}
\end{keypoints}

%
%

%
%


\begin{abstract}
Machine learning models for the global atmosphere that are capable of producing stable, multi-year simulations of Earth’s climate have recently been developed. However, the ability of these ML models to generalize beyond the training distribution remains an open question. In this study, we evaluate the climate response of several state-of-the-art ML models (ACE2-ERA5, NeuralGCM, and cBottle) to a uniform sea surface temperature warming, a widely used benchmark for evaluating climate change. We assess each ML model’s performance relative to a physics-based general circulation model (NOAA's Geophysical Fluid Dynamics Laboratory AM4) across key diagnostics, including surface air temperature, precipitation, temperature and wind profiles, and top-of-atmosphere radiation. While the ML models reproduce key aspects of the physical model response, particularly the response of precipitation, some exhibit notable departures from robust physical responses, including radiative responses and land region warming. Our results highlight the promise and current limitations of ML models for climate change applications and suggest that further improvements are needed for robust out-of-sample generalization.
\end{abstract}

\section*{Plain Language Summary}
Recent machine learning (ML) models can simulate Earth’s atmosphere for many years without becoming unstable, raising the possibility that they could be used to study climate change. A key unanswered question, however, is whether these models can reliably predict climate responses under conditions they were not trained on, such as a warmer world. In this study, we test several leading ML-based climate models by applying a simple and widely used experiment: uniformly warming the ocean surface. We compare how these ML models respond to warming against a well-established physics-based climate model. We examine changes in temperature, rainfall, atmospheric circulation, and radiation. We find that the ML models successfully capture some major features of the climate response, especially changes in precipitation. However, they also show important differences from the physics-based model, particularly in how radiation responds to warming and how temperatures change over land areas. These differences suggest that, while ML climate models are promising, they are not yet fully reliable for predicting climate change outside the conditions they were trained on. Overall, our results show that machine learning has strong potential for climate modeling, but further development is needed before these models can be confidently used for climate change projections.
%
%

\section{Introduction}
A variety of machine learning (ML) approaches have been developed for weather and climate science. Substantial progress has been made in ML-based weather prediction, where forecast skill scores are now competitive with, and in some cases surpass, those of state-of-the-art numerical weather prediction models used by leading forecasting centers \cite{lam2023learning,bi2023accurate,rasp2024weatherbench,keisler2022forecasting}. A key advantage of ML models is their computational efficiency, with costs that are a fraction of those required by traditional, physical numerical weather prediction models.

Extending ML approaches to climate-timescale atmospheric modeling, however, presents additional challenges. These include the need for stable inference over periods much longer than the training data (e.g., centuries rather than decades) and the ability to generalize to climate states outside the training distribution (e.g., levels of warming not yet observed but expected in the coming century), which could potentially be addressed by incorporating physical knowledge into ML models \cite{beucler2024climate}. Despite these challenges, ML may be central to the next generation of multiscale climate modeling for mitigation and adaptation strategies \cite{eyring2024ai}. At a minimum, it is one component of the emerging future of climate modeling \cite{bordoni2025futures}.


An important appeal of ML methods is the potential to exploit the reduced computational cost to refine climate models without requiring higher spatial resolution \cite{eyring2024ai}. ML approaches can help capture essential Earth system processes and feedbacks, while remaining efficient enough to generate the large ensembles needed to study internal variability, extremes, and climate attribution. Recently, \cite{bracco2025machine} reviewed advances in applying ML to the physics of climate. Overall, ML offers the potential to reduce computational cost, improve accuracy (e.g., via observational calibration), and enable the generation of very large ensembles for climate simulations. 

One ML approach to atmospheric modeling is a hybrid modeling approach, where ML is used in combination with physical components. This can take the form of replacing parameterizations with ML schemes or learning the tendencies from a higher resolution model to use at lower resolution \cite{bretherton22}. An end-member of this approach is to retain the dynamical core and explicitly simulate large-scale atmospheric flows and replace all parameterizations of physical processes (in particular, the atmospheric water cycle and radiative transfer) with learned physics \cite{kochkov2024neural}.

Alternatively, ML architectures can be trained to emulate the whole general circulation model, which has been done for both the atmosphere and ocean \cite{watt2023ace,dheeshjith2025samudra,clark2024ace2}. To achieve stable whole model emulators, conservation blocks in the architecture ensure key physical properties such as conservation of atmospheric mass or water vapor are respected in the training process \cite{watt2025ace2,chapman2025camulator}. \add{Training data itself may not precisely conserve energy or mass, as is the case for atmospheric reanalyses due to data assimilation tendencies, so an ML model potentially learns unphysical relationships without hard constraints from conservation blocks.}

In climate change research, a fundamental benchmark is the atmospheric response to uniform sea surface temperature (SST) warming \cite{cess90, eyring16, merlis24}. This simple sensitivity test provides a useful standard for evaluating a model’s climate sensitivity. Diagnostics such as top-of-atmosphere (TOA) radiation balance, mean and extreme precipitation, vertical temperature profiles, and large-scale circulation responses capture essential physical processes and sensitivities underlying climate projections. This is a valuable component test of atmospheric models as it is simple to set up, rapidly equilibrates, and has well understood changes that capture the leading-order behavior of the atmospheric response to warming. 

In this study, we perform a model intercomparison and evaluate the climate responses of three ML-based models—cBottle \cite{brenowitz2025climate}, ACE2-ERA5 \cite{watt2025ace2}, and NeuralGCM \cite{yuval2024neural}—to a uniform +2 K SST warming scenario. A limited evaluation of SST warming was presented in the ACE2 and NeuralGCM documentation papers \cite{watt2025ace2,yuval2024neural}. \add{Recently,} \citeA{chen2026hierarchical} \add{presented results of NeuralGCM in response to a uniform SST warming of 4 K. We include cBottle and ACE2-ERA5 to expand the variety of ML model types. In addition,} \citeA{rucker2026benchmarking} \add{focuses on the ability of ML models to simulate trends over historical periods and showed varied success at regional levels. In this study, we focus more on the response to uniform SST warming instead of the trend. Moreover, the AI Weather and Climate Model Intercomparison Project (AIMIP) has recently been proposed to enable a more comprehensive evaluation of ML models} \cite{henn2026aimip}.
Brief descriptions of these ML models are provided in the Methods section. Their performance is compared with the Geophysical Fluid Dynamics Laboratory's AM4 \cite{zhao2018gfdlam4part1,zhao2018gfdlam4part2}, which serves as a physically based reference. Our analysis focuses on land surface warming patterns, hydrological changes, zonal-mean temperature and wind structures, and radiative responses under idealized warming. The output variables of the public releases of the ML models are not exhaustive, so models are omitted from the analysis as necessary. Through this intercomparison, we assess the strengths and limitations of current ML models and their potential role in future climate applications.

\section{Methods}
\subsection{ML Models}
In this study, we examine three global ML atmospheric models. They are:
\begin{enumerate}
\item Climate in a Bottle, cBottle: cBottle is a generative diffusion model developed by NVIDIA \cite{brenowitz2025climate}. It consists of two complementary diffusion models: a coarse-generation model and a super-resolution model.  While the super-resolution model version is useful for mesoscale weather and climate features, we exclusively evaluate the coarse model here. The coarse-generation model produces outputs at a resolution of approximately 100 km, comparable to that of typical CMIP6-era climate models \cite{eyring16}. This makes it a suitable tool for conducting multi-year climate emulation and investigating the large-scale response to SST perturbations. It is trained on ERA5 over the years 1980-2017. Atmospheric variables, such as wind and temperature, are represented on 8 pressure levels. 
cBottle takes monthly mean SST as input and generates hourly climate outputs, including the SST itself. It is distinctive in that it is not auto-regressive: the state of the atmosphere does not depend on its history. In this study, we conduct both the control and +2 K simulations for 10 years each, using the default set of diagnostic variables listed in \cite{brenowitz2025climate}. The monthly SST used in the control simulation is taken from the AMIP simulations that are part of the CMIP6 project \cite{eyring16}, with climatological values computed for the period 1981–2014. Note that cBottle allows control of stochastic variability through a user-defined “seed” parameter. In the simulations presented in the manuscript, this parameter was not specified. However, we performed two additional pairs of control and +2 K experiments with the seed set to 42 and 52, respectively, and found the results to be consistent, indicating that the results are robust and not sensitive to the seed setting.

\item ACE2-ERA5: ACE2-ERA5 is an auto-regressive ML model that uses Spherical Fourier Neural Operator (SFNO) architecture and is trained to emulate the global atmosphere by learning from the ERA5 reanalysis dataset \cite{watt2025ace2}. \remove{It employs a two-stage architecture consisting of a convolutional encoder and a Transformer-based predictor with mass and water vapor conservation blocks to generate rollouts of key atmospheric variables, such as temperature, winds, and precipitation.} \add{ACE2's SFNO architecture is combined with a physical corrector module that enforces exact conservation of global dry air mass and column-integrated atmospheric moisture. It is an autoregressive emulator that takes 6-hour time steps to produce rollouts of key atmospheric variables, including temperature, winds, humidity, and precipitation.} Input and output variables for ACE2 can be found in supporting information in \cite{watt2025ace2}. ACE2-ERA5 operates on a regular latitude–longitude grid with 1$^\circ$ spatial resolution and 6-hourly temporal resolution, comparable to reanalysis products and suitable for weather-to-climate scale applications. There are 8 vertical layers in ACE2.
The model can produce stable multi-year simulations when initialized with real-world atmospheric states and forced with observed sea surface temperatures (SSTs). 
ACE2-ERA5 uses SST and sea ice as forcing data to do the emulation and advances the atmospheric state over 6-hour steps. The 6-hourly SST  and sea-ice forcing data is a coarsened version of the ERA5 reanalysis. We conduct both the control and +2 K simulations for 10 years each, using the default set of diagnostic variables listed in \cite{watt2025ace2}. Climatological values for the control simulation are computed over the period 1981–2014. Note that CO$_2$ concentration is also included as an input to ACE2-ERA5; however, we keep the CO$_2$ value identical in both the control and +2 K simulations. ACE2-ERA5 requires an initial condition, with available years including 1940, 1950, 1979, 2001, and 2020. Here, we use the initial condition from the year 2001 for both the control and +2 K experiments, with CO$_2$ concentration is about 370 ppm. Note that the overall performance is insensitive to the choice of initial conditions. \add{This protocol separates the co-variation of SST and CO$_2$ in ACE2-ERA5's training dataset.}

\item NeuralGCM is a hybrid ML model, combining a differentiable dynamical core  with neural network parameterizations of all physical processes \cite{kochkov2024neural}. As such, NeuralGCM is not a whole model emulator like  ACE2-ERA5 and cBottle. NeuralGCM has been shown to produce skillful medium-range forecasts and multi-year climate simulations. When forced with prescribed SSTs, NeuralGCM achieves stable decadal-scale simulations at resolutions of 140–280 km, capturing essential climate features such as seasonal cycles, tropical cyclone statistics, and large-scale circulation patterns. 
We save daily outputs for NeuralGCM, which has a time-step governed by the dynamical core (i.e., comparable to GCMs). An updated version of NeuralGCM includes precipitation as an output variable \cite{yuval2026neural}, and we use this version of NeuralGCM to examine the response of precipitation along with other variables to uniform SST warming. \add{This version is trained on satellite-based precipitation observations, so it may learn different relationships between precipitation and other variables compared to models trained solely on ERA5.} Note that NeuralGCM can occasionally go unstable. We found this to be the case when using the climatological SST that we used for the ACE2 roll-out with 2 K of warming. For NeuralGCM, we used the ERA5 SST climatology over 2000-2020 (to better match its 2001-2018 training period), which was stable for 10 years for both the Control and +2 K simulations. All results presented here are based on these 10-year integrations. We also found that the public release of this precipitation version of NeuralGCM had long-term drifts that eventually went unstable when using NVidia H100, H200, and L40S Graphical Processing Units and consequently performed these rollouts on Google's Colab platform with v6e-1 Tensor Processing Unit hardware. 
\end{enumerate}

In all cases, the input to the ML models was an SST climatology, rather than the interannually varying SST fields typically used in the Atmospheric Model Intercomparison Protocol (AMIP). A preliminary investigation suggests this choice does not substantially affect the results and allows for the shorter 10 year comparisons presented here.



\subsection{Atmospheric General Circulation Model}
We conduct simulations using the AM4 general circulation model, developed by the Geophysical Fluid Dynamics Laboratory (GFDL). Model descriptions of AM4 can be found in \cite{zhao2018gfdlam4part1,zhao2018gfdlam4part2}.
Briefly, AM4 features a cubed-sphere dynamical core with 96 × 96 grid cells per face, corresponding to an approximate horizontal resolution of 100 km. For analysis and comparison, model output is post-processed and re-gridded onto a regular latitude–longitude grid with 180 meridional points and 288 zonal points, yielding a resolution of 1.0° latitude × 1.25° longitude. The model includes comprehensive physical parameterizations, including moist convection, radiation, cloud physics, and boundary layer turbulence. The control simulation is forced with monthly climatological sea surface temperatures (SSTs) and sea ice concentrations averaged over the period 1981–2014. To evaluate the model’s response to warming, we perform a uniform SST warming experiment with a +2 K anomaly applied globally over the ocean surface. Comparisons of ML models and AM4 are shown in Table \ref{tab:model_summary}.

While we present the results of a single atmospheric GCM, we note that the amip protocol with uniform SST warming is an experiment in the Cloud Feedbacks Model Intercomparison protocol \cite{webb17}. There, the perturbation is 4 K, and we opted to evaluate a smaller magnitude perturbation so that the ML models were closer to their training data. The response of GCMs is  typically approximately linear in the perturbation amplitude and is appears that ACE2-ERA5's surface air temperature response is linear in SST perturbations \cite{watt2025ace2}.

\subsection{Observations}
Global Precipitation Climatology Project, GPCP, v3.2 \cite{huffman2023new} is used to evaluate the mean precipitation climatology. Radiation at the top of the atmosphere from CERES EBAF Edition 4.2.1 \cite{NASA/LARC/SD/ASDC2024CERES} is used to evaluate the models’ energy budget. Monthly ERA5 reanalysis data \cite{hersbach20} is used for the surface air temperature.

\begin{sidewaystable}
\centering
\caption{Comparison of ML models and AM4}
\label{tab:model_summary}
\vspace{10pt} 
\begin{tabular}{l l l l p{7cm}}
\toprule
\textbf{Model} & \textbf{Spatial Res.} & \textbf{Temp. Res.} & \textbf{Type} & \textbf{Output Variables Analyzed} \\
\midrule
cBottle [coarse] & 100 km & hourly & Generative ML & tas, pr, column water vapor, u, v, t, OLR, SW up \\
\addlinespace 
ACE2-ERA5 & 100 km & 6-hourly & Autoregressive ML & tas, pr, evaporation (surface latent heat flux), column water vapor, u, v, t, OLR, SW up\\
\addlinespace
NeuralGCM & 280 km & customizable & Hybrid ML-GCM & t1000, pr, evaporation, specific humidity (to compute column water vapor), u, v, t \\
\addlinespace
AM4 & 100 km & customizable & Physics-based GCM & full atmospheric diagnostics \\
\bottomrule
\end{tabular}
\end{sidewaystable}

\section{Results}

\subsection{Surface Air Temperature}

\begin{figure}
    \centering
    \includegraphics[width=\linewidth]{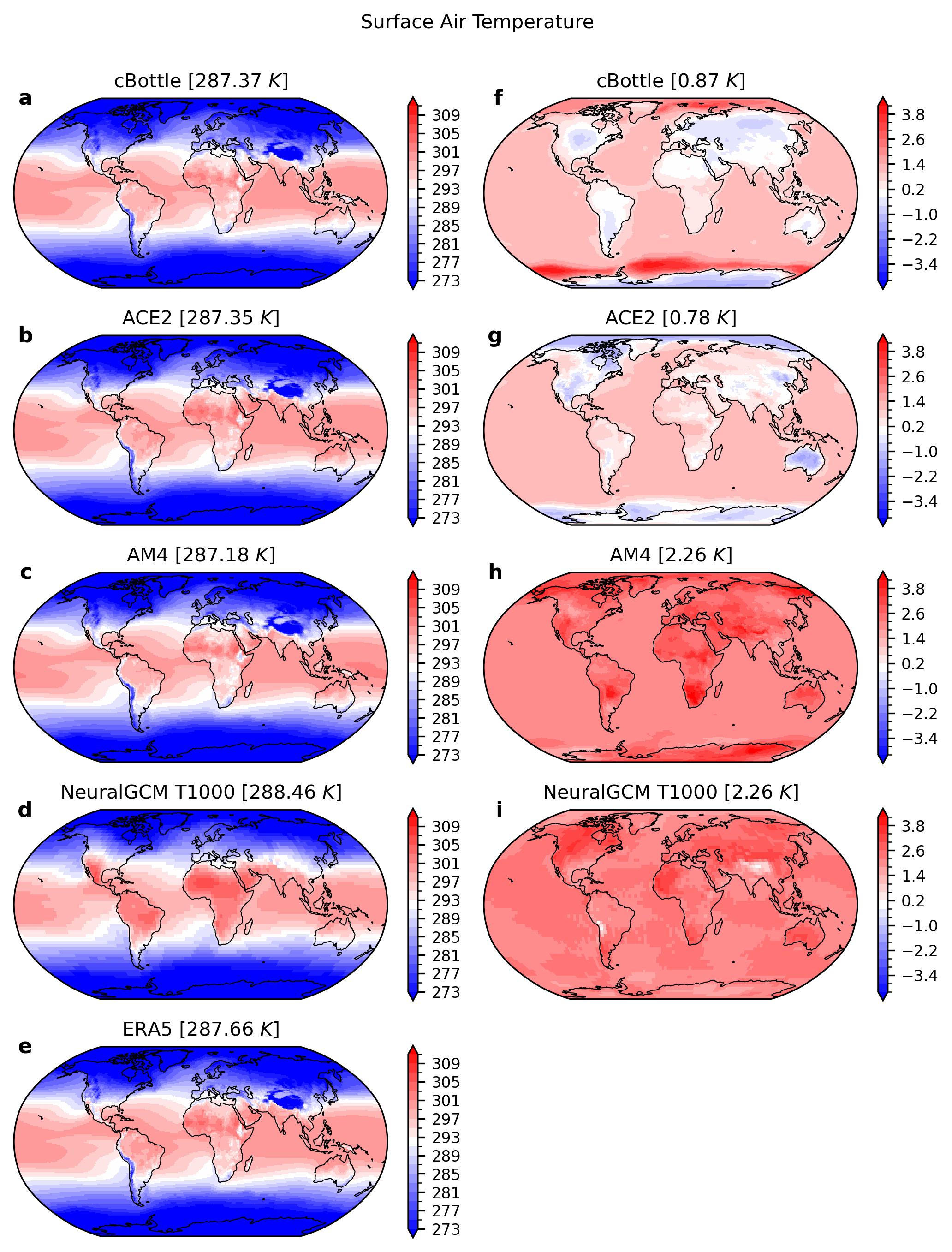}
    \caption{Annual mean surface air temperature for (left) the climatology and (right) the response to a uniform +2 K SST perturbation for cBottle, ACE2, AM4, and NeuralGCM from top to bottom. ERA5 is included in panel (e) for reference of the mean state. Note that NeuralGCM does not directly output surface air temperature. Here we use temperature at 1000 hPa to illustrate the temperature pattern near the surface. \add{The numbers shown next to each panel title indicate the global mean values.}}
    \label{fig:t2m}
\end{figure}

Overall, all ML models reproduce the mean climatology of surface air temperature reasonably well compared with AM4 and the European Centre for Medium-Range Weather Forecasts' ERA5 renalysis (Fig.~\ref{fig:t2m}a-e). The public release of NeuralGCM does not provide surface air temperature directly; instead, we use the 1000 hPa air temperature as a proxy, so its
high-altitude regions are not comparable, as the temperature there includes subsurface extrapolation.
\add{(A version of NeuralGCM with additional output features, including surface air temperature, can be found in Fig.~1 of} \citeA{henn2026aimip}.)
The global-mean surface air temperatures, indicated above the maps in Fig.~\ref{fig:t2m}, are close to that of ERA5 for all models.

\begin{figure}
    \centering
    \includegraphics[width=\linewidth]{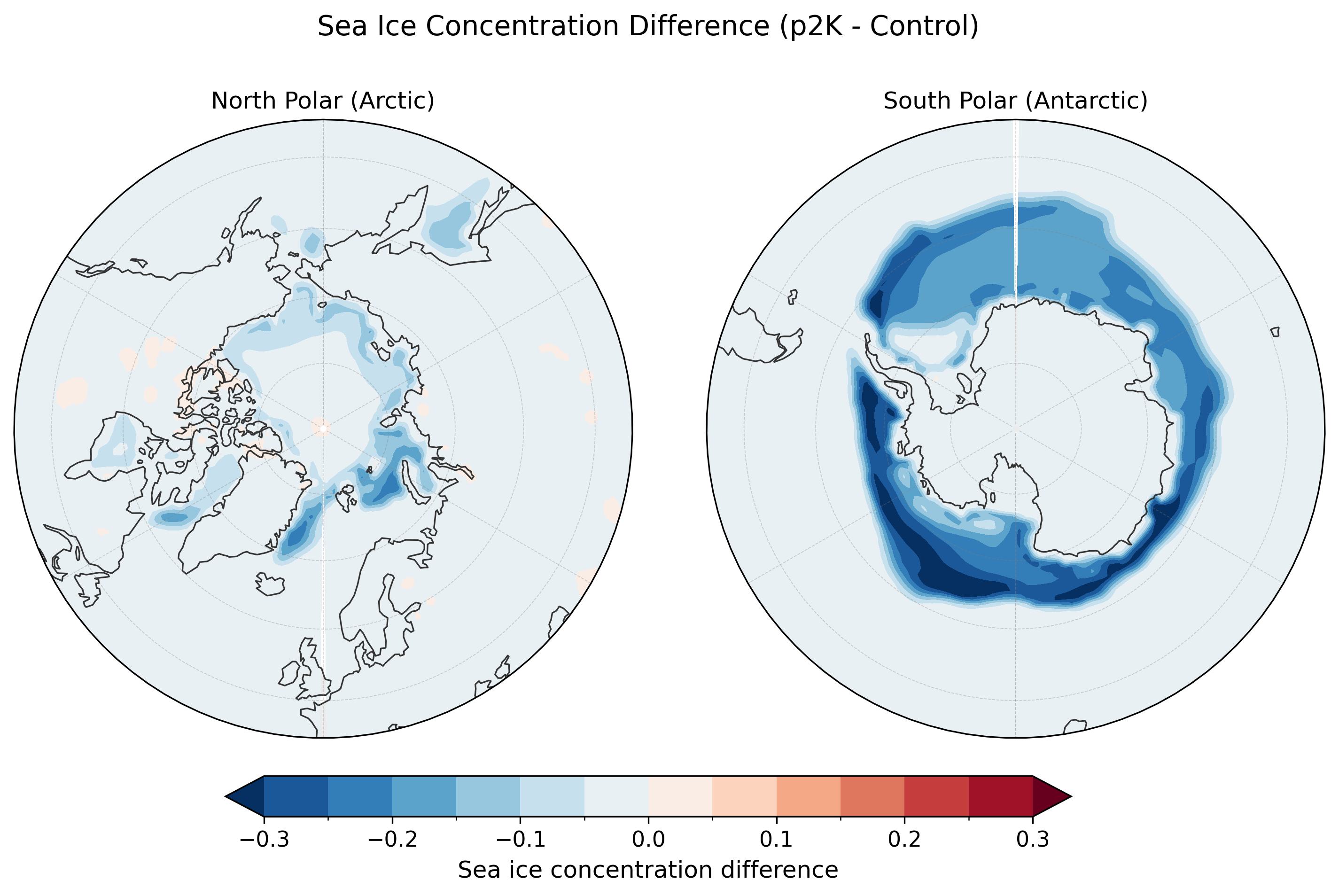}
    \caption{Annual mean sea ice response to a uniform +2 K SST perturbation in cBottle, shown for the Arctic (left) and Antarctic (right).}
    \label{fig:sic}
\end{figure}

In response to uniform SST warming, AM4 exhibits pronounced land warming (Fig. \ref{fig:t2m}h), which is a well-understood result of enhanced surface warming over dry surfaces because of the relationship between relative humidity near the surface and the atmospheric lapse rate \cite{joshi08,byrne13a,byrne13b}. This amplified warming over land is underestimated in cBottle (Fig. \ref{fig:t2m}f). Nonetheless, cBottle simulates amplified polar warming, likely due to sea-ice loss induced by SST warming \remove{(not shown)}. \add{The sea ice reduction is larger in the Southern Ocean than in the Arctic.} (\ref{fig:sic}) Notably, sea ice is an output in cBottle, making it slightly different from conventional atmosphere-only simulations with prescribed SST and sea ice. In contrast, ACE2 shows a surface air temperature response over land that is a mix of warming and cooling (Fig. \ref{fig:t2m}g)---a limitation also noted by \cite{watt2025ace2}. \add{The global-mean land surface temperature difference is 0.12 K for cBottle, 0.26 K for ACE2, and 2.67 K for AM4. }Enhanced land warming is well captured in NeuralGCM compared to AM4, although the magnitude of polar warming is underestimated (Fig. \ref{fig:t2m}i). This is suggestive of an important role for NeuralGCM's dynamical core in capturing the land--sea mechanism \cite{byrne13a} that posits weak temperature gradients aloft and differences in lower-tropospheric lapse rate set the amplitude of land enhanced warming. While ACE2-ERA5 does not have enhanced land warming, ACE2 does have enhanced land warming when trained across physical model simulations of multiple climates \cite{clark2024ace2}.

The land-enhanced warming in AM4 and other General Circulation Models (GCMs) means that the global-mean surface air temperature exceeds that of the imposed SST perturbation. This is the case for the $1000 \, \mathrm{hPa}$ temperature in NeuralGCM, but both ACE2 and cBottle have less global-mean surface air temperature change than the imposed SST increase. Beyond the land response, the ocean region response can also contribute. In particular, we find that cBottle has SST and surface air temperature increases over the ocean that are less than the imposed $2 \, \mathrm{K}$ perturbation (Fig. \ref{fig:t2m}f). \add{A similar behavior is also found in ACE2} (Fig. \ref{fig:t2m}g). \add{These results are consistent with those from the AIMIP results} \cite{henn2026aimip}. Analogous cold biases have been documented in ML-based weather forecasts for years beyond the end of the training period, as there has been additional warming that the models do not generalize to \cite{rackow24,landsberg25}.

\subsection{Precipitation}

\begin{figure}
    \centering
    \includegraphics[width=\linewidth]{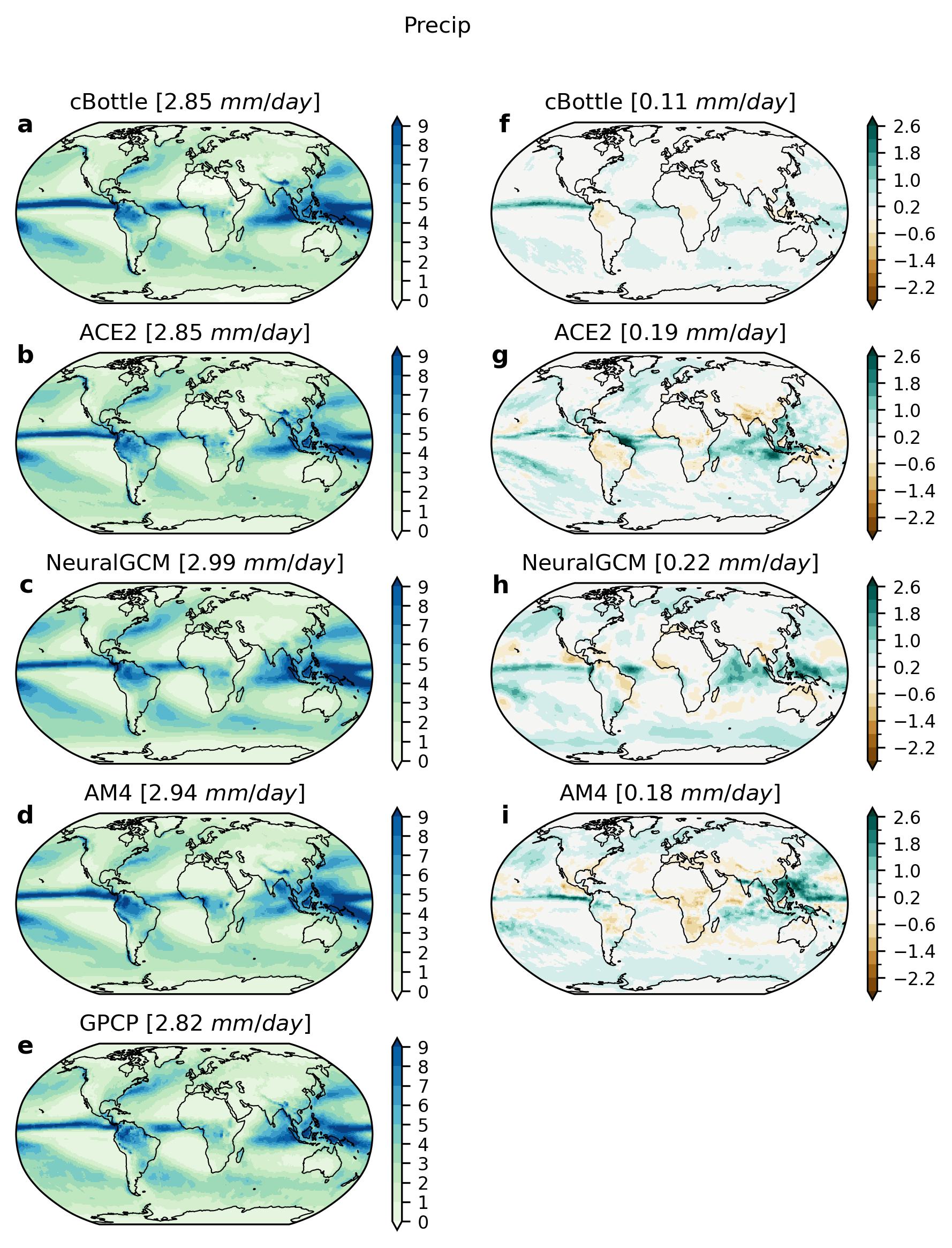}
    \caption{Annual mean precipitation for (left) the climatology and (right) the response to a uniform +2 K SST perturbation for cBottle, ACE2, NeuralGCM and AM4, from top to bottom. GPCP is included in panel (e) for reference.}
    \label{fig:pr}
\end{figure}

The mean climatology of precipitation is reasonably well captured by all models (Fig.~\ref{fig:pr}a-e). The global-mean precipitation simulated by cBottle, ACE2, and AM4 agrees closely with observed estimates from the GPCP dataset, while NeuralGCM produces slightly higher values (Fig.~\ref{fig:pr}). 

Under uniform SST warming, all models simulate an increase in global-mean precipitation, with the strongest regional increases in the tropics (Fig.~\ref{fig:pr}f-h). The magnitude of the global-mean precipitation change is quantitatively similar to the GCM shown here and the robustly simulated magnitude of a $\approx 3\% \, \mathrm{K}^{-1}$ increase that is driven by increased radiative cooling \cite{allen02, jeevanjee18}.
One regional discrepancy is cBottle's weak increase in northern hemisphere mid-latitude.
There are regions in the low latitudes, particularly over land and the subtropical oceans, where the GCM response is a decrease in precipitation and the ML models show similar behavior to varying degrees. This is broadly consistent with the thermodynamic, ``wet-get-wetter, dry-get-drier'' response \cite{held06}, which we evaluate next. 

\begin{figure}
    \centering
    \includegraphics[width=\linewidth,height=0.8\textheight,keepaspectratio]{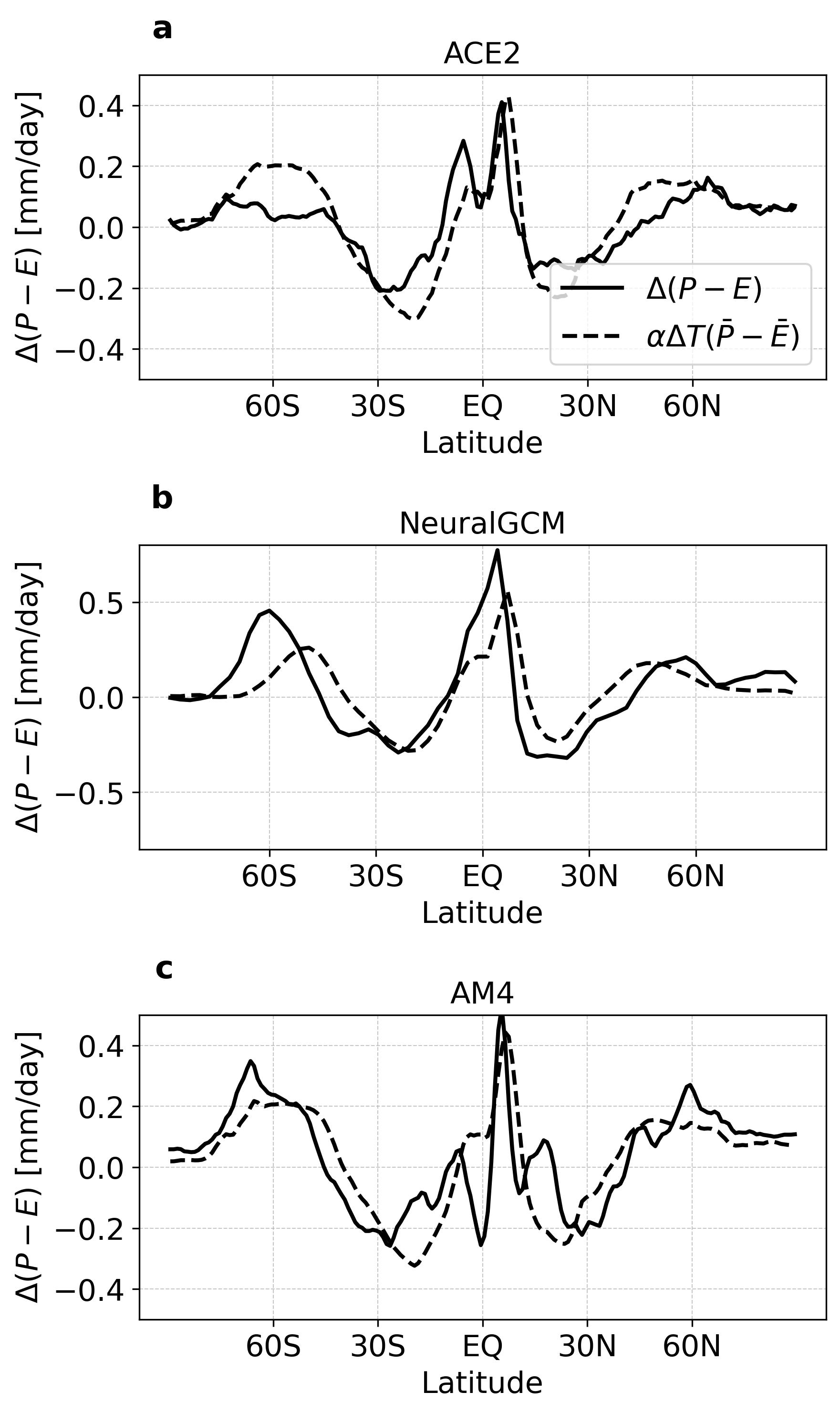}
    \caption{The zonal mean change of precipitation minus evaporation for ACE2, NeuralGCM and AM4 from top to bottom. The dashed lines in each panel is the thermodynamic component,approximated as $\alpha \Delta T(\bar{P}-\bar{E})$ with $\Delta T=2K$ and $\alpha = 0.07 \, \mathrm{K^{-1}}$.}
    \label{fig:pminuse}
\end{figure}

Precipitation minus evaporation (P–E) is governed by the vertically integrated water vapor transport, which has thermodynamic increases under warming because the saturation vapor pressure of air increases with temperature according to the Clausius-Clapeyron relation \cite{held06}. We show changes in P-E in Fig.~\ref{fig:pminuse}. The evaporation field is unavailable in cBottle, so it is excluded. ACE2, NeuralGCM, and AM4 all exhibit the canonical ``wet-get-wetter, dry-get-drier'' response \cite{held06}, with increased P–E along the Intertropical Convergence Zone (ITCZ), reduced P–E in subtropical subsidence regions, and increased P-E in the extratropics (Fig. \ref{fig:pminuse}). In addition, these models show deviations from the thermodynamic response approximated by $\alpha \Delta T(\bar{P}-\bar{E})$ (dashed lines in Fig.~\ref{fig:pminuse}), including a poleward expansion of subtropical dry zones. This is an impressive degree of agreement with physical models for a non-trivial aspect of the response to warming.

It is interesting and surprising that these ML models generalize the precipitation rate to warmed atmospheres. This raises the question of what aspects of today's climate, as encapsulated in the ERA5 training dataset, give rise to this behavior. While it is a challenge to interpret large neural networks (e.g., ACE2 has 450 million parameters), we think it is useful to consider intervariable relationships that are known to be important in physical models. First, the specific humidity $q$ of historical atmospheric states is constrained by the Clausius-Clapeyron relation that governs the saturation vapor pressure: $q \leq q^*(T,p) $, where $q^*(T,p)$ is the temperature and pressure-dependent saturation specific humidity. A thermodynamic increase in humidity with warming is embedded in today's atmospheric state. Second, a relationship between humidity and precipitation rates would allow for the warmer temperatures to translate to precipitation changes.

\begin{figure}
    \centering
    \includegraphics[width=\linewidth]{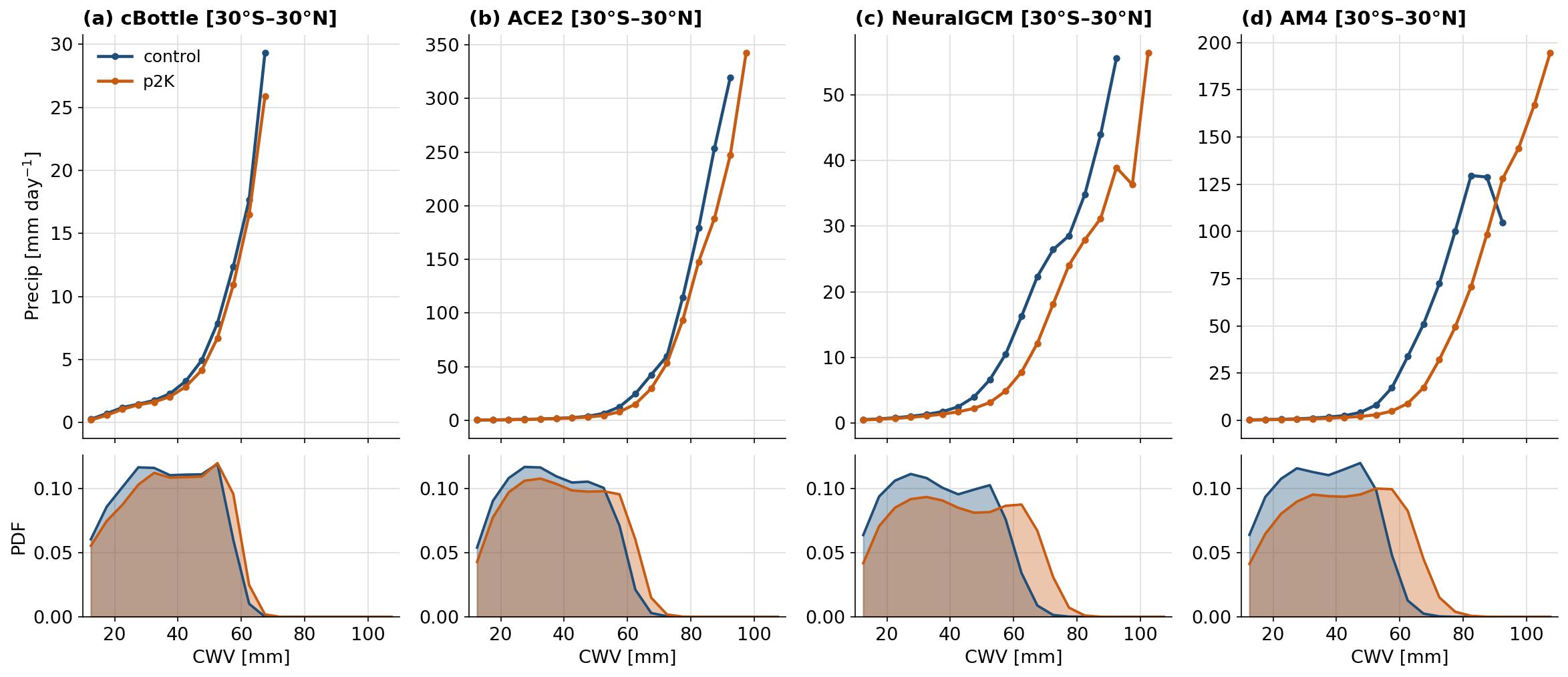}
    \caption{Daily mean precipitation versus binned column water vapor in the tropics (30°S–30°N) for (a) cBottle, (b) ACE2, (c) NeuralGCM, and (d) AM4 for the control simulation (blue) and the +2K simulation (orange). The bottom row shows the probability distribution function of the sample size (i.e., the frequency) for each column water vapor bin.}

    \label{fig:cwv}
\end{figure}

With this interpretative lens, we analyze the relationship between daily mean precipitation and column water vapor (CWV) in the tropics \cite{bretherton04b,kuo2020convective, neelin2022precipitation}. There is typically weak precipitation until a sufficiently high value of CWV $\approx 50 \, \mathrm{mm}$, where there is a transition to a sensitive increase in precipitation with CWV \cite{bretherton04b}. All of the models capture some form of this climatological precipitation--CWV relationship (Fig. \ref{fig:cwv}). With SST warming, the precipitation ``onset'' value of CWV increases. Physically, this is expected because the saturation deficit increases with warming if the relative humidity is approximately unchanged. At the highest values of CWV, particularly those that exceed the maxima of the control climate, the precipitation associated with the highest CWV bins becomes stronger. All ML models have some shift in the physically expected direction, although the degree to which the onset CWV shifts to higher values varies between the ML models, with the $\approx 10 \, \mathrm{mm}$ of NeuralGCM most comparable to AM4. At the highest CWV bin, all models have an increase in the precipitation rate, though all ML models have a weaker increase than AM4. 
\add{The frequency of a given value of CWV can also change with warming. In physical models, relative humidity is approximately unchanged with warming, so the CWV distribution shifts to higher values. Taking AM4 as a reference, we see that NeuralGCM has comparable shifts in the CWV frequency with warming} (Fig. \ref{fig:cwv}, bottom row), \add{ while ACE2 and cBottle have much weaker shifts toward higher CWV with warming. Given these differences in the frequency of CWV values, one cannot directly infer the behavior of precipitation extremes of a given percentile from} Fig. \ref{fig:cwv}, so we examine that next.
In summary, it appears that there are compensating errors in the changes in the precipitation-CWV relationship in the perturbed climates: the modest shift in onset of strong precipitation tending to increase the mean precipitation more than physical models, but the \add{CWV distribution shift toward higher values is weaker in the ML models and there is a} weaker increase in precipitation at the highest CWV that mutes the increase. 


\begin{figure}
    \centering
    \includegraphics[width=\linewidth]{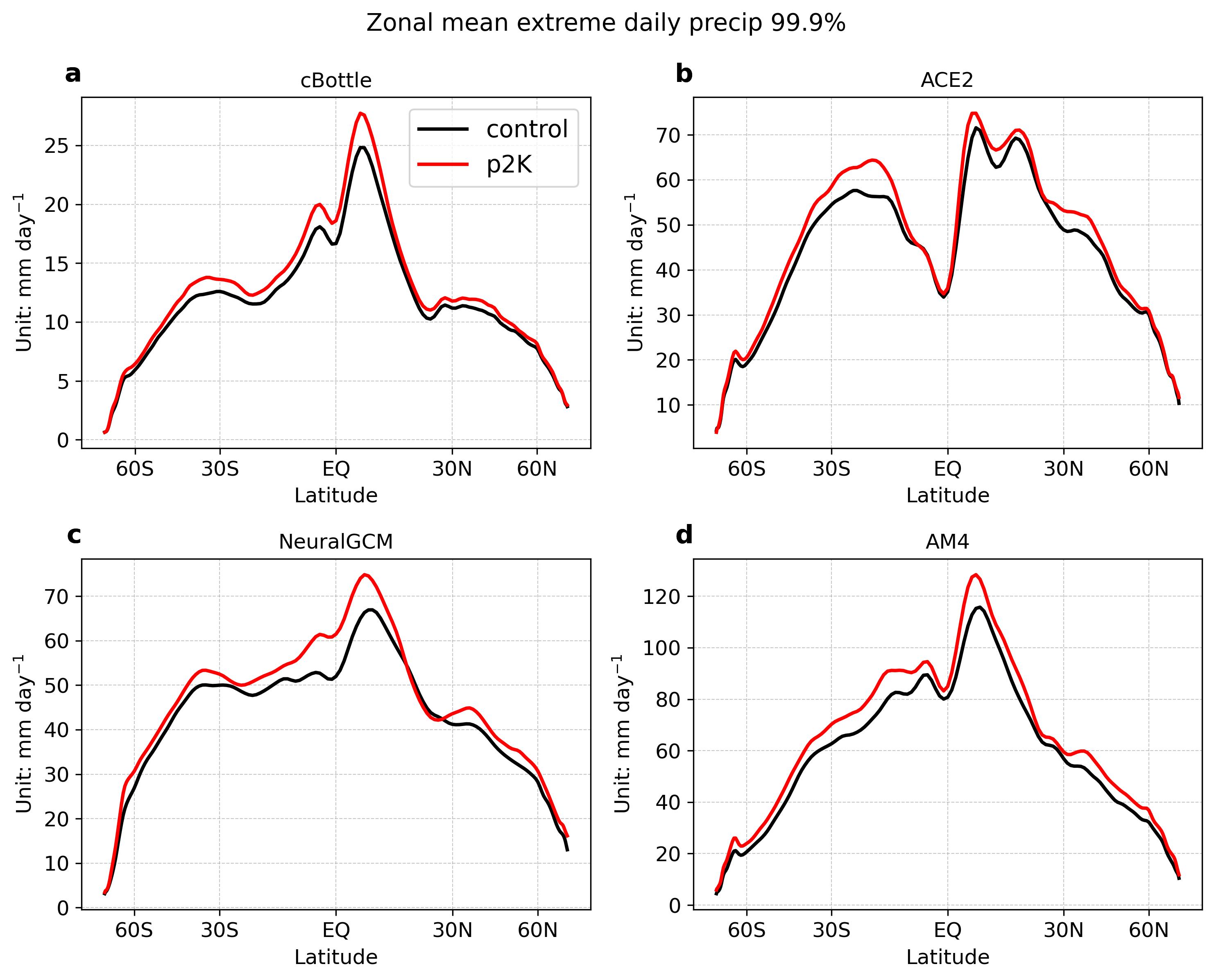}
    \caption{Zonal mean 99.9th percentile of daily precipitation for (a) cBottle, (b) ACE2, (c) NeuralGCM, and (d) AM4. Results are shown for the control simulation (black) and the +2K simulation (red). }

    \label{fig:extreme}
\end{figure}

We also examine extreme precipitation, using the 99.9th percentile of daily precipitation at each grid point, following \cite{OGorman09a,OGorman09b}. The zonal mean of this metric is shown in Fig.~\ref{fig:extreme}. There are substantial differences in the magnitude of the control precipitation extremes, with cBottle having the smallest values. \add{Examining cBottle's hourly output reveals extreme precipitation rates in-line with ERA5, so the low values for daily precipitation extremes appear to result from inadequately representing temporal correlations found in nature's and physical models' precipitation rates.} ACE2's magnitude of $\approx 50 \, \mathrm{mm \, day^{-1}}$ in the tropics is similar to the ERA5 data on which it was trained. The ML models capture the equator-to-pole contrast in the precipitation extremes, though ACE2 has locally suppressed extremes near the equator.

Under uniform SST warming, extreme precipitation intensifies across all models (Fig. \ref{fig:extreme}). 
The structure of the extremes in the warmed climate largely follows that of the control, with increases in the tropics (30°S to 30°N) that are $\approx 4.5 \% \ \mathrm{K}^{-1}$ for cBottle, $\approx 3.1 \% \ \mathrm{K}^{-1}$ for ACE2, $\approx 3.8 \% \ \mathrm{K}^{-1}$ for NeuralGCM, and $\approx 4.6 \% \ \mathrm{K}^{-1}$ for AM4. These changes are smaller than that estimated by the Clausius-Clapeyron relationship of saturation vapor pressure of about 6-7 $\% \ \mathrm{K}^{-1}$. \add{In physical models, circulation changes can also influence how extreme precipitation responds to warming} \cite<e.g.,>[]{OGorman09a}, \add{so it is interesting that the ML models have a consistent sub-Clausius-Clapeyron rate of change that would typically be interpreted as a weakening of the vertical velocity of the extremes.}

\subsection{Atmospheric Temperature and Circulation Response}

\begin{figure}
    \centering
    \includegraphics[width=\linewidth]{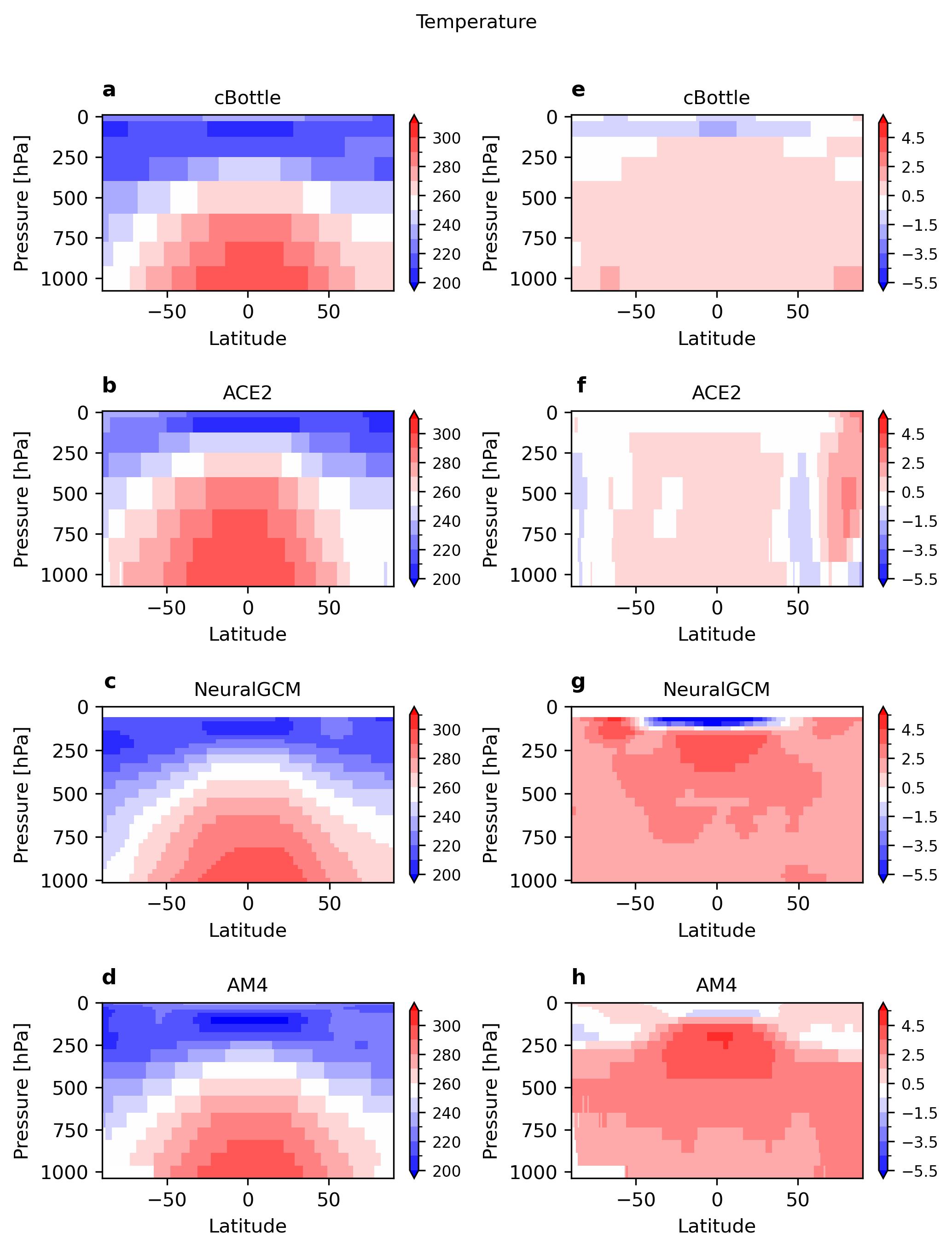}
    \caption{Annual- and zonal-mean (left) control simulation temperature and (right) temperature response to +2K SST warming of (a,e) cBottle, (b,f) ACE2, (c,g) NeuralGCM, and (d,h) AM4. Note that pressure levels below 40 hPa are masked out for NeuralGCM because the model top corresponds to $p/p_s = 0.03$ \cite{kochkov2024neural}.}
    \label{fig:temp}
\end{figure}

The climatological zonal-mean temperature profiles are shown in the left column of Fig. \ref{fig:temp}. The temperature distribution is reasonable across all models, with a tropical tropospheric temperature maximum that decreases poleward.

In response to SST warming, AM4 exhibits pronounced upper-tropospheric warming in the tropics (Fig. \ref{fig:temp}h), consistent with moist adiabatic adjustment driven by increased latent heat release from deep convection \cite{santer05}.
This amplified warming is well reproduced by NeuralGCM (Fig. \ref{fig:temp}g) but is substantially weaker in both cBottle and ACE2 (Figs. \ref{fig:temp}e and f), where the upper-tropospheric warming is similar to the near surface-temperature response in the tropics. \citeA{rucker2026benchmarking} \add{showed that ML models can capture historical trends reasonably well. While this is not unexpected given that these recent trends are in the training dataset, the response to uniform SST warming shown here represents a more stringent out-of-sample test and ACE2's behavior is quite different.}

The closer agreement between NeuralGCM and AM4 may reflect the role of a dynamical core in both models, which is absent in cBottle and ACE2, or may reflect differences in its training approach. NeuralGCM simulates notable stratospheric cooling under uniform SST warming (Fig. \ref{fig:temp}g), whereas the other models show little to no such signal. Stratospheric cooling in warmed climates is dominated by temperature adjustments to carbon dioxide forcing (rather than being proportional to the SST warming), and ACE2 captures this when trained on physical model simulations that have a slab ocean boundary condition and perturbed carbon dioxide concentration \cite{clark2024ace2}.

\begin{figure}
    \centering
    \includegraphics[width=\linewidth]{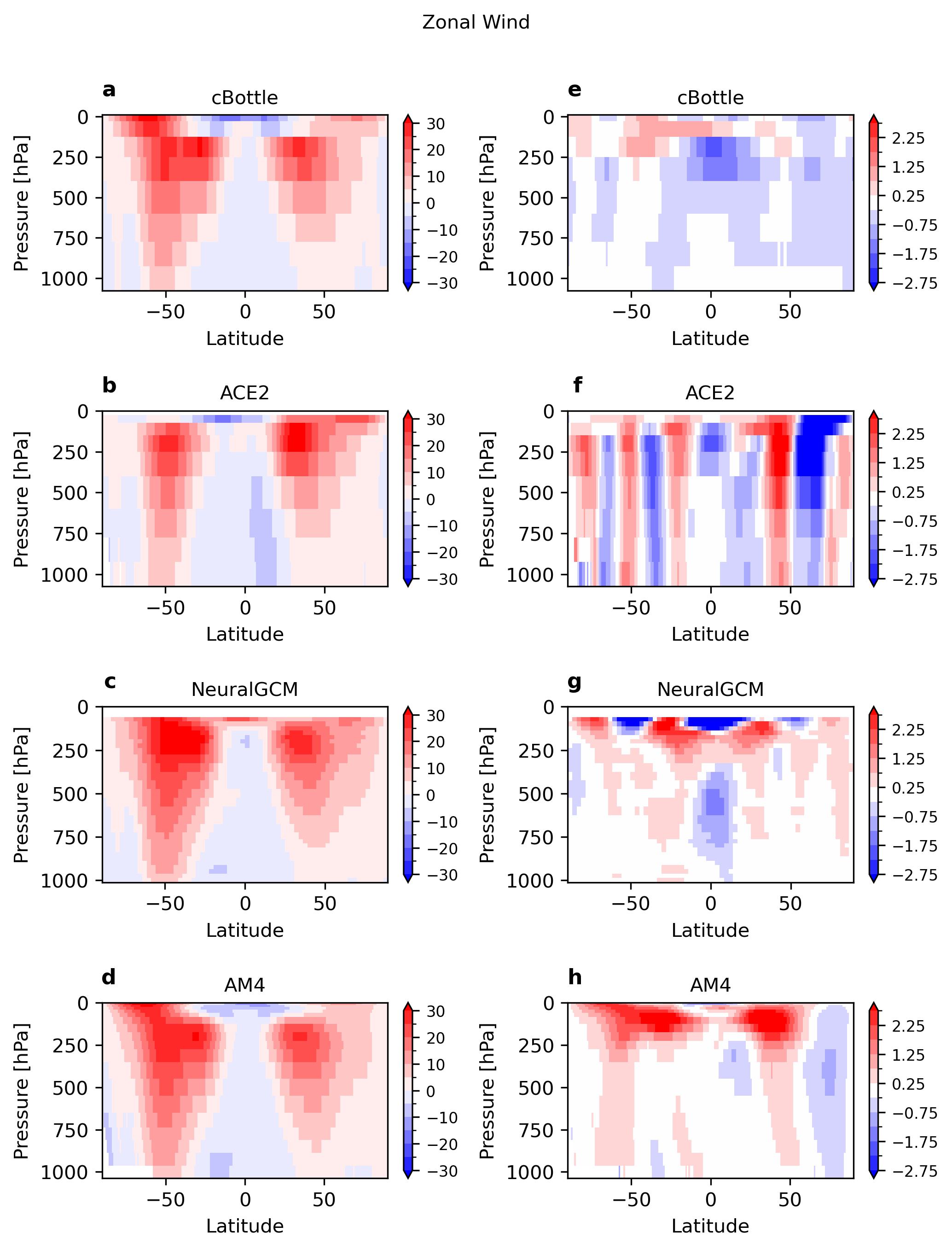}
    \caption{Annual- and zonal-mean (left) control simulation zonal wind and (right) zonal wind response to +2K SST warming of (a,e) cBottle, (b,f) ACE2, (c,g) NeuralGCM, and (d,h) AM4. As in Fig. \ref{fig:temp}, pressure levels below 40 hPa are masked out for NeuralGCM.}

    \label{fig:wind}
\end{figure}

Similar to temperature, the climatological zonal-mean zonal wind is realistically captured by all models, featuring upper-tropospheric westerly jets in both hemispheres and tropical easterlies. One climatological bias is the width of the westerly jet in the southern hemisphere extratropics in ACE2, which occupies a narrower range of latitudes (Fig.~\ref{fig:wind}d). A more detailed assessment of the variability of the present-day atmospheric circulation in ACE2 and NeuralGCM can be found in \cite{baxter25}.

The zonal-mean zonal wind response to SST warming differs substantially across models (Figure \ref{fig:wind}). Overall, cBottle and ACE2 display zonal wind responses of both signs (Fig. \ref{fig:wind}e and f), with meridional structure on the scale of $\approx 10^\circ$ that do not bear clear relationships to the climatological winds. In contrast, NeuralGCM and AM4 simulate a strengthening of upper-tropospheric westerlies (Fig. \ref{fig:wind}g and h), consistent with their warming patterns aloft (Fig. \ref{fig:temp}g and h) and the thermal wind balance. AM4 has  poleward shifts in the surface westerlies, a longstanding result of GCM simulations of climate change \cite{kushner01}. NeuralGCM and cBottle have weak changes in the surface zonal wind, while ACE2 has the narrower scale changes extending to the surface. NeuralGCM also produces anomalous stratospheric easterlies (Fig. \ref{fig:wind}g), which are absent in AM4 and are linked via thermal wind to the exaggerated stratospheric cooling in NeuralGCM (Fig. \ref{fig:temp}g).

\begin{figure}
    \centering
    \includegraphics[width=\linewidth]{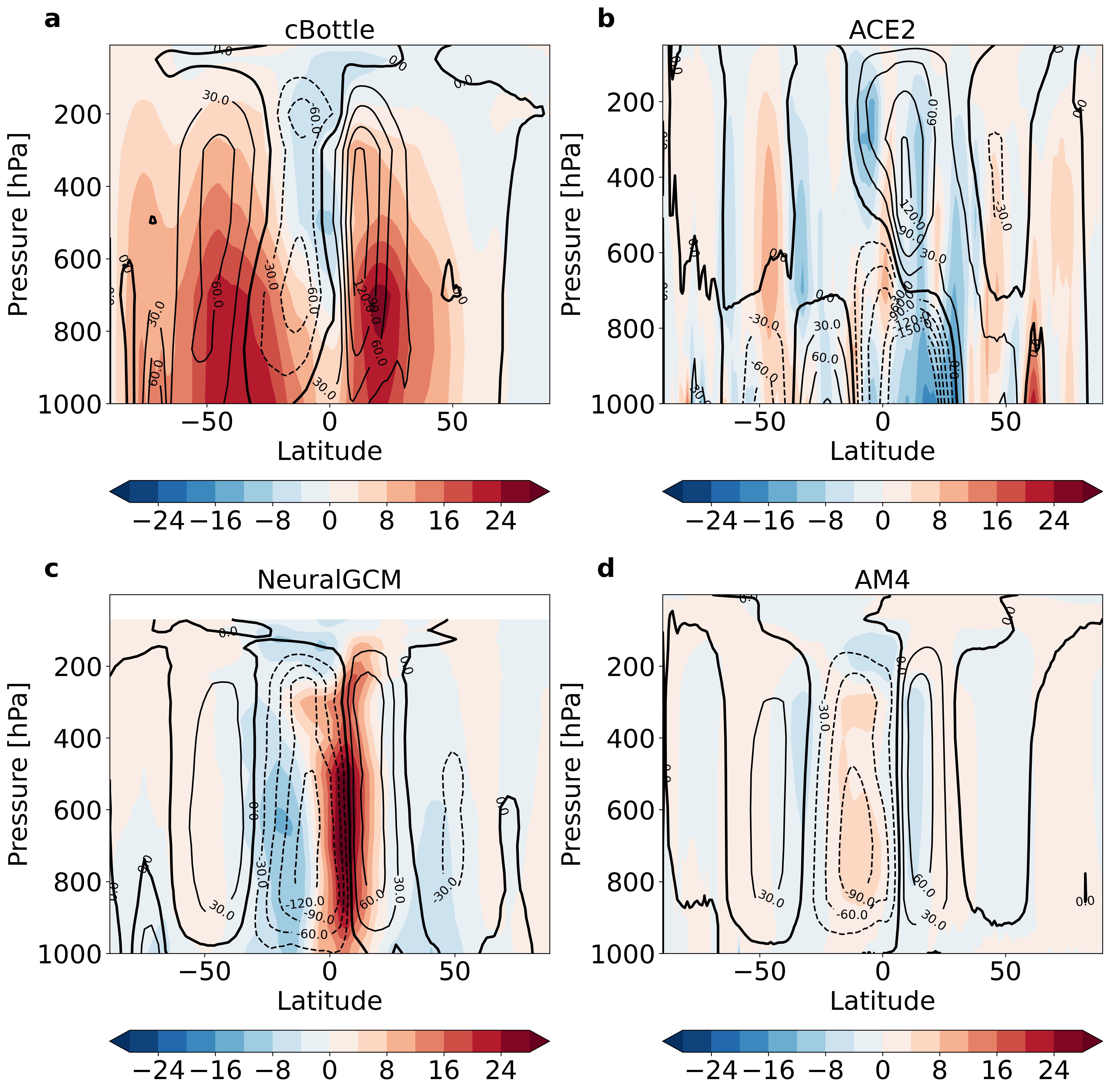}
    \caption{Annual- and zonal-mean meridional streamfunction for (a) cBottle, (b) ACE2, (c) NeuralGCM, and (d) AM4. Contours denote the climatological mean state, and shading shows the response to +2 K SST warming (p2K minus control). Contours range from 150 to 150 at intervals of 30, with the zero contour shown as a thick line. Units are $\times 10^{9} \ \mathrm{kg \ s^{-1}}$. As in Fig. \ref{fig:temp}, pressure levels below 40 hPa are masked out for NeuralGCM.} 

    \label{fig:streamf}
\end{figure}

Figure~\ref{fig:streamf} shows zonal-mean meridional streamfunction. Unlike the zonal wind, the mean state of the meridional streamfunction differs substantially across models. In cBottle, the northern hemisphere Hadley cell is stronger than its southern hemisphere counterpart (Fig. \ref{fig:streamf}a), while ACE2 shows clear deficiencies in simulating the Hadley circulation (Fig. \ref{fig:streamf}b).
This discrepancy between the climatology of the meridional wind in the ML models and Earth's climatology suggests that the mean meridional wind is a more difficult variable to train, perhaps owing to its smaller variance in the tropics compared to the extratropics. 

Under SST warming, cBottle has a strengthening of the northern hemisphere Hadley cell and a vertically varying response in the southern hemisphere, with near-surface weakening and upper tropospheric strengthening (Fig. \ref{fig:streamf}a). ACE2 shows mixed signals, with changes in the sign of the streamfunction response over small meridional scales (Fig. \ref{fig:streamf}b). NeuralGCM has an equatorward shift of the ITCZ (Fig. \ref{fig:streamf}c) and the P-E change is physically consistent with this (Fig. \ref{fig:pminuse}b). 
AM4 exhibits a weakening of the Hadley cell in both hemispheres (Fig. \ref{fig:streamf}d) by about $\approx 1\%$, which is in line with most GCMs projecting a weakening of the Hadley cell, although the magnitude is uncertain, ranging from $0\%\ \mathrm{K}^{-1}$ to $4\%\ \mathrm{K}^{-1}$ \cite{vecchi2007global,d2017factors,lionello2024hadley}.

\subsection{Radiation}
\begin{figure}
    \centering
    \includegraphics[width=\linewidth]{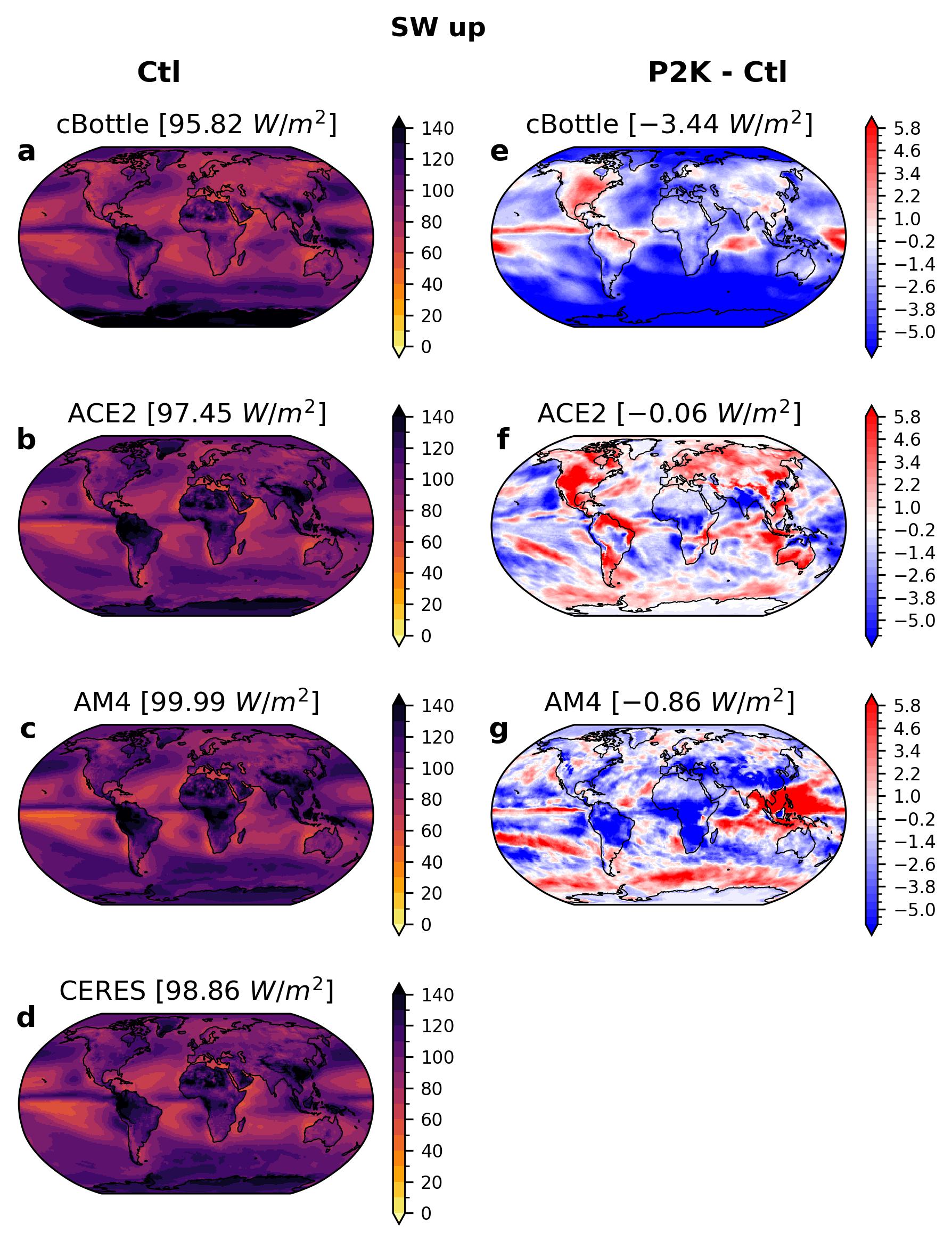}
    \caption{Annual-mean upward SW radiation at TOA for (left) the control simulation and (right) the response to +2K SST perturbation for cBottle, ACE2, and AM4 from top to bottom. CERES is shown in panel (d) for reference.}

    \label{fig:sw}
\end{figure}

TOA radiation is another key diagnostic of atmospheric models. NeuralGCM is excluded from this analysis because it does not provide TOA radiation fields. We first examine the mean climatology of upward shortwave (SW) radiation at TOA. cBottle slightly underestimates the global-mean upward SW flux, while ACE2 and AM4 produce global means that are closer to CERES observations (Fig. \ref{fig:sw}). 

The response of upward SW radiation to SST warming varies substantially across the three models. In cBottle, the upward SW flux is strongly reduced almost everywhere except in the deep tropics (Fig. \ref{fig:sw}e). In contrast, ACE2 and AM4 show much smaller decreases in global-mean upward SW radiation—less than $1 \, \mathrm{W \, m^{-2}}$ (Fig. \ref{fig:sw}f and g). A decrease in the global-mean reflected solar radiation is typical among GCMs \cite{merlis24}. The spatial pattern of responses differs between ACE2 and AM4, with ACE2 exhibiting positive anomalies over most land regions (except Africa) and AM4 having predominantly negative anomalies over land. We note that the SW cloud response to warming is uncertain (i.e., there is substantial intermodel spread between GCMs).

\begin{figure}
    \centering
    \includegraphics[width=\linewidth]{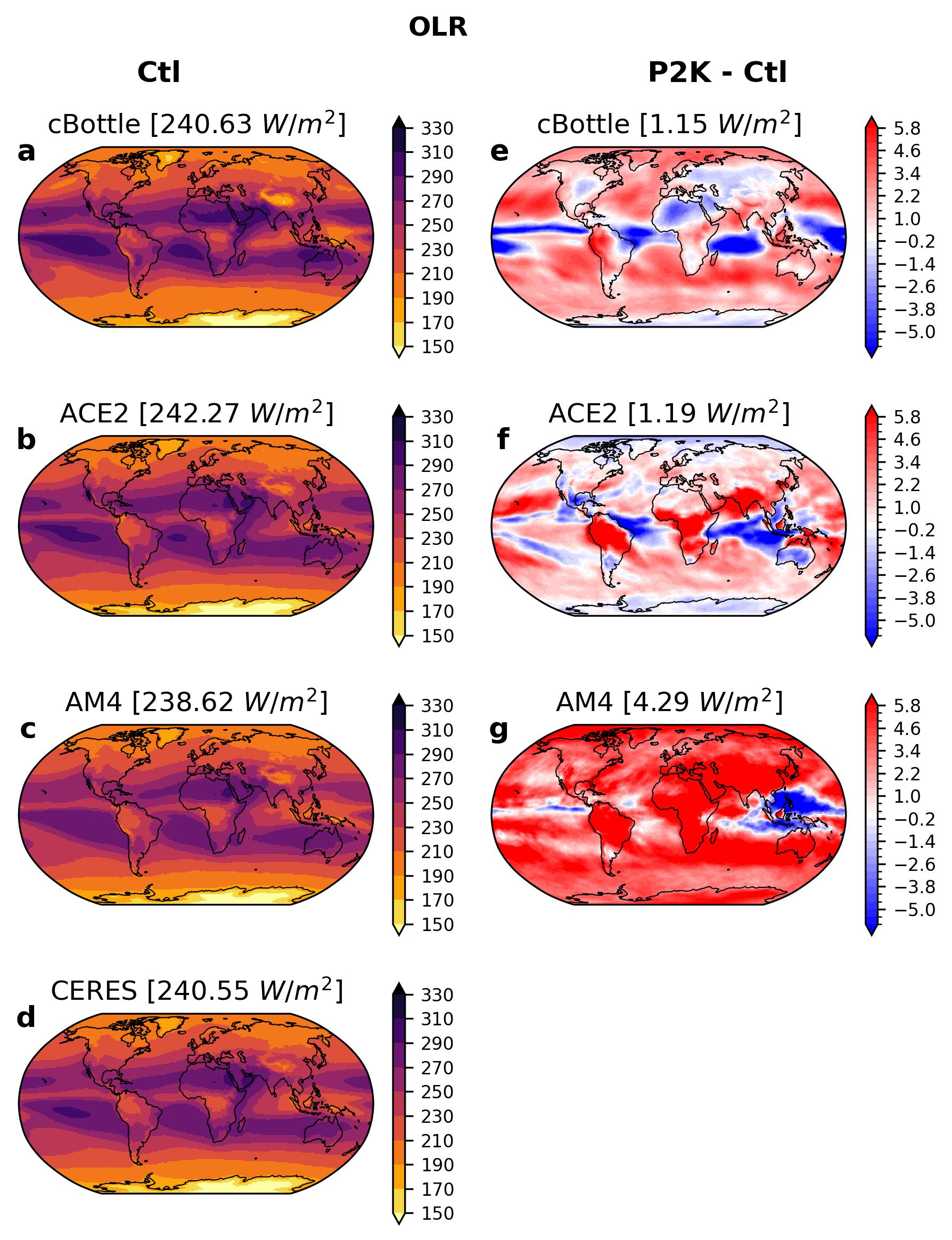}
    \caption{Annual-mean OLR at TOA for (left) the control simulation and (right) the response to +2K SST perturbation for cBottle, ACE2, and AM4 from top to bottom. CERES is shown in panel (d) for reference.}

    \label{fig:lw}
\end{figure}

For outgoing longwave radiation (OLR), the mean climatology in all models agrees reasonably well with observations, with global mean values near $240 \, \mathrm{W \, m^{-2}}$ (Fig. \ref{fig:lw}). The basic regional features of this climatology are represented in all models: the OLR is lower in the deep tropics, where there are high clouds and high humidity, compared to the drier subtropics, and there is an equator-to-pole reduction in OLR as the emission decreases with mean temperature.

Under SST warming, the global-mean OLR increases in all models, indicating enhanced longwave cooling. However, the magnitude of this cooling tendency is weaker in both cBottle and ACE2 compared to AM4 (Fig. \ref{fig:lw}). As a result, the combined SW and LW responses in cBottle lead to a net positive energy imbalance. This is physically unrealistic, as the implied feedback parameter would produce a runaway warming if the SST were allowed to respond consistent with the energy budget changes.

\begin{figure}
    \centering
    \includegraphics[width=\linewidth]{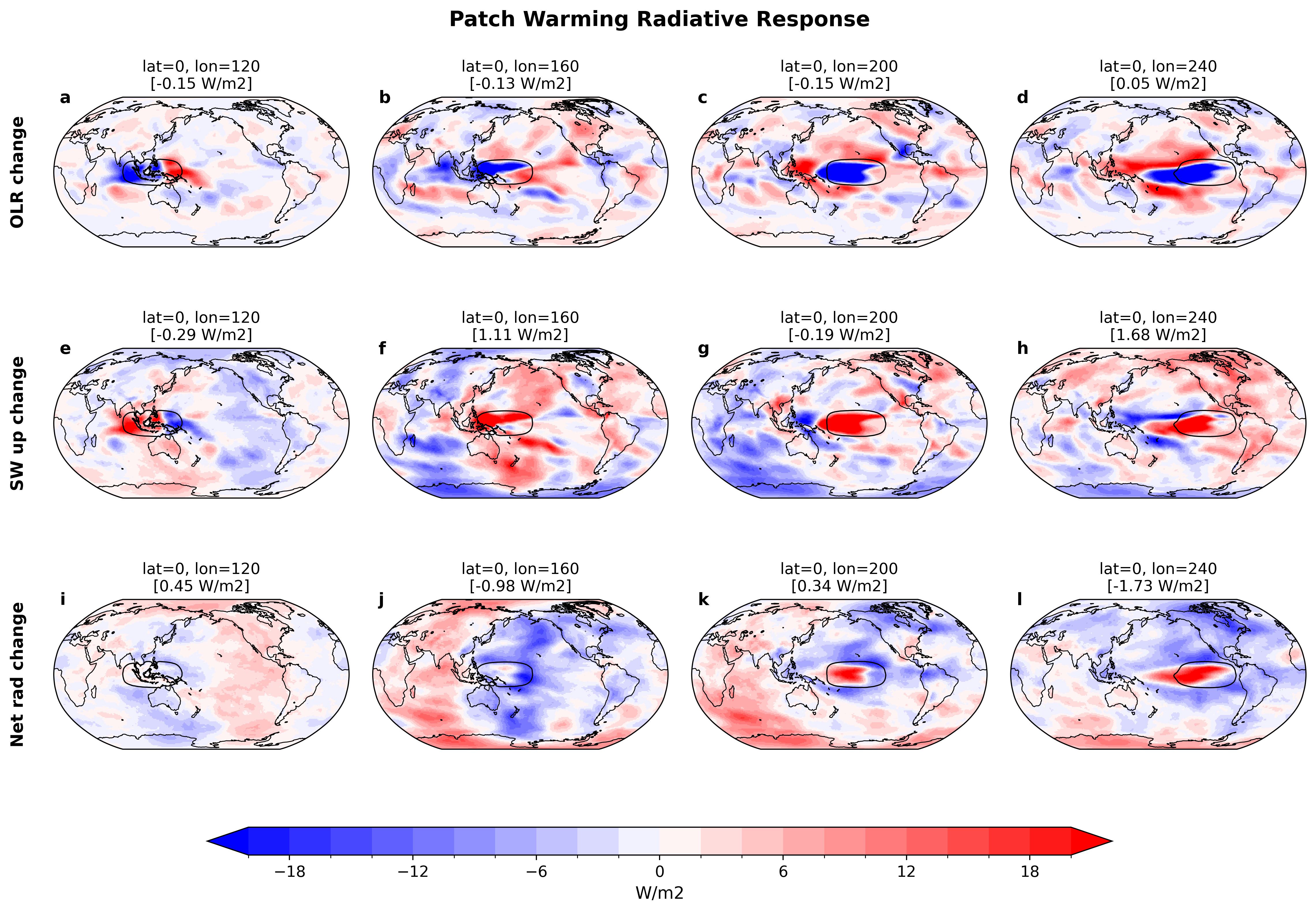}
    \caption{Annual-mean radiative flux responses to patch warming experiments, 
    shown as anomalies relative to the control. Columns correspond to patch 
    centroids at the equator ($\mathrm{lat}=0^\circ$) with $\mathrm{lon}=120^\circ$, 
    $160^\circ$, $200^\circ$, and $240^\circ$. Rows show anomalies in OLR (top), 
    reflected shortwave radiation (SW$\uparrow$, middle), and net TOA radiation 
    (bottom). Black contours mark the SST anomaly threshold ($0.1$~K), and 
    bracketed values in each panel indicate area-weighted global-mean anomalies 
    (W~m$^{-2}$).}
    \label{fig:patch}
\end{figure}

To shed light on the extent to which this behavior is particular to the uniform SST perturbation or is found generically in cBottle, we also conducted additional experiments in which only localized (patch) SST warming anomalies were imposed, following the Green’s function–style experiments previously performed with AM4 \cite{zhang2023using,bloch2024green} and for ACE2 \cite{wu2025applying,vonloon25}. \remove{In this case, cBottle exhibits a negative global-mean net TOA radiation response to the SST patch warming, indicating a stable feedback (not shown).} \add{We find that cBottle produces both positive and negative global-mean net TOA radiation responses as the SST warming patch is shifted from the tropical western Pacific to the eastern Pacific} (Fig. \ref{fig:patch}).\add{ This indicates that cBottle is capable of representing stable feedback, although its response to patch SST warming does not necessarily resemble that of conventional physics-based models. For example, the two western-most patches are typically have the most stable radiatively restoring (i.e., biggest decrease in net radiation) in GCMs, while the western-most cBottle patch here has an increase in net radiation. In contrast to the tropical west Pacific, the tropical east Pacific is a region that is locally unstable in physical models with a decrease in upward SW from a low cloud reduction. In cBottle, there is a local increase in upward SW in the east Pacific and the global-mean upward SW is an increase } (Fig. \ref{fig:patch}). A recent study also applied Green’s-function–style experiments to the ACE2 model and found results that are physically consistent with those from traditional physics-based models, albeit with quantitative differences \cite{wu2025applying}.
\remove{The contrasting behaviors between the uniform and patched SST warming experiments suggest that the idealized uniform SST warming case represents a truly out-of-sample prediction, where the model is unable to extrapolate reliably, whereas the regional patch experiments have perturbations that are closer to the model’s training distribution (e.g., historical El Ni\~no-Southern Oscillation events).} \add{One might hope that the behavior of the regional patch experiments is more credible given that the perturbations are closer to the model's training distribution (e.g., historical El Ni\~no-Southern Oscillation events), while the uniform SST warming case is a truly out-of-sample prediction. Previous work using ACE2 has shown promising results for these perturbation experiments, while the cBottle results here depart substantially from physical model behavior.}

\add{Returning to the uniform SST warming perturbation,} ACE2 simulates a decrease in TOA net radiation under warming, \add{although} its magnitude is much weaker than in AM4. In other words, AM4 exhibits a more stable climate feedback parameter than ACE2, driven primarily by larger OLR response. Normalizing the change in TOA flux by the surface-air temperature change brings ACE2 and AM4's feedback parameter closer together because of the muted surface warming in ACE2.

\begin{figure}
    \centering
    \includegraphics[width=\linewidth]{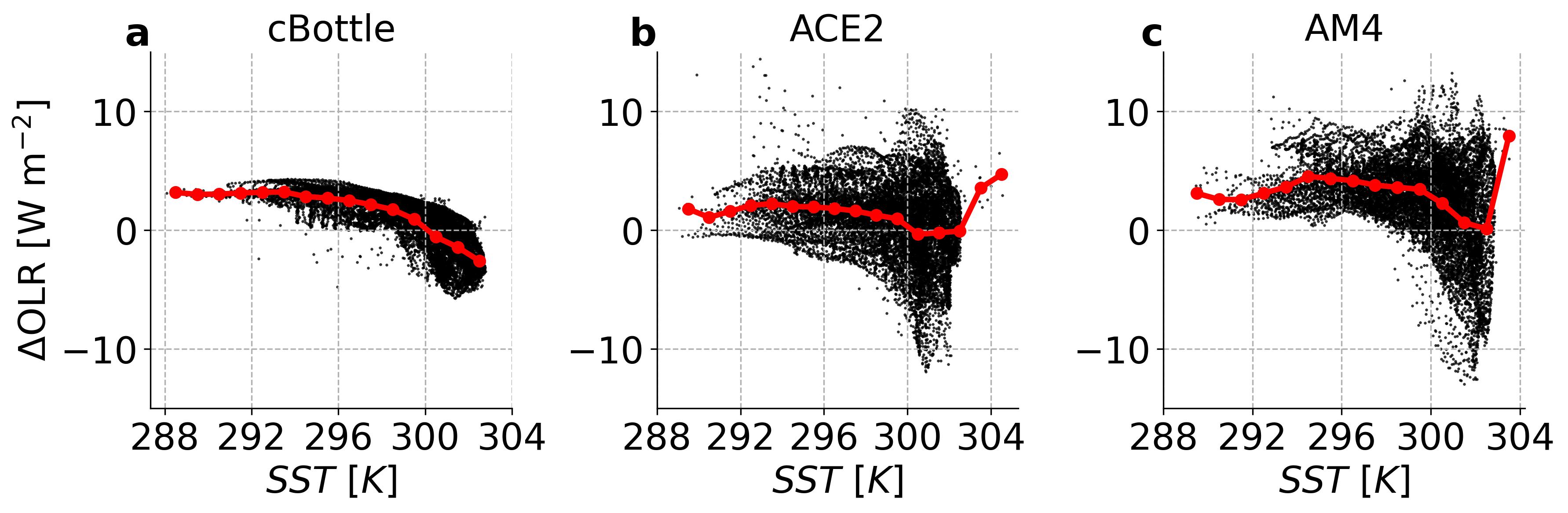}
    \caption{The response of OLR to +2K SST warming vs the control SST at each grid point (black dots) for (a) cBottle, (b) ACE2, and (c) AM4. The red solid line in each panel denotes the mean OLR response within SST bins from 288 to 305 K, using 1 K intervals.}
    \label{fig:scatter}
\end{figure}

The spatial pattern of the change in OLR has a common reduction in OLR over tropical deep convective regions across the models. In a physical model, one can interpret this as reflecting an expansion of high cloud area. It could also indicate higher cloud tops under warmer SSTs, though typically the cloud top temperatures are unchanged with warming \cite{hartmann02}. Scatter plots of OLR change versus climatological SST in the tropics (Fig. \ref{fig:scatter}) show that negative OLR responses are consistently associated with high-SST regions, characteristic of active deep convection. Above 300 K, cBottle’s OLR response is predominantly negative, whereas ACE2 and AM4 display a mixture of positive and negative responses.



\section{Summary and Discussion}

We compared the mean state and climate responses to uniform SST warming in three atmospheric ML models—cBottle, ACE2, and NeuralGCM, using the physics-based AM4 GCM as a reference. Observations were also used to evaluate the mean state of each model. To first order, all ML models reproduce the observed climatology reasonably well, including air temperature, precipitation, circulation, and radiation fields. This agreement is not surprising, given that ERA5 reanalysis data were used in their training. Two aspects of the control climatology that differ across models are the Eulerian-mean mass transport (i.e., the mean meridional wind) and the extreme precipitation metric (99.9th percentile of daily precipitation rate) assessed here. 

The response to uniform SST warming provides a straightforward measure of a model’s sensitivity to external forcing. For ML models, this experiment serves as an out-of-sample test. We find the ML models exhibit substantially different responses to SST warming, including some discrepancies from well understood physical responses. 

For surface air temperature, cBottle and ACE2 underestimate the overall magnitude of warming. Differences in model structure account for cBottle's  polar warming (associated with its determination of sea ice as an output of the inference). In contrast, ACE2 has high latitude cooling. Neither cBottle nor ACE2 has enhanced land surface warming, which is a robustly simulated GCM response. This issue is ameliorated when ACE2 is trained on physical model simulations of perturbed climates \cite{clark2024ace2}. NeuralGCM's lowest atmospheric temperature response is enhanced over land regions, consistent with the known physical response. 

The precipitation response to warming is well captured by the models assessed here. All exhibit an increase in global-mean precipitation and regional changes, such as the largest increase in the deep tropics, which are characteristic of GCM-simulated changes. Both ACE2 and NeuralGCM capture the canonical ‘wet-get-wetter, dry-get-drier’ pattern seen in AM4 and other physics-based models, resulting from thermodynamic increases in atmospheric water vapor with temperature \cite{held06}. Because the relationships among temperature, humidity, and precipitation are inherently embedded in the training data (e.g., ERA5), the ML models can reproduce the overall pattern of precipitation changes driven by increases in humidity and temperature under uniform SST warming. The strong coupling between precipitation and humidity is illustrated in Fig. \ref{fig:cwv}, where all models show a rapid increase in precipitation once column water vapor exceeds approximately 50 mm. In addition, the ML models also have expansions of subtropical dry zones that are known to result from wind changes in physical models. Last, all models simulate an intensification of extreme precipitation under SST warming, though the magnitude of this response varies quantitatively.

For atmospheric temperature responses, NeuralGCM and AM4 both show amplified warming in the tropical upper troposphere, whereas cBottle and ACE2 fail to reproduce this long-standing feature of climate model simulations. \add{Interestingly, ACE2 does produce trends in tropical temperature in recent decades that are similar to its ERA5 training dataset} \cite{rucker2026benchmarking}, \add{but that structure is not present in the uniform SST warming case.}  Similarly, intensified upper-tropospheric westerlies are simulated by NeuralGCM and AM4, but not by cBottle and ACE2. The physical consistency---or lack thereof---in the relationship between the zonal wind and temperature change is evidence of the dynamical core in constraining the atmospheric response to SST warming, as it ensures dynamical relationships such as thermal wind balances are respected. 

The meridional streamfunction changes in the ML models are much larger in magnitude than in AM4, and the spatial patterns in cBottle and ACE2 show little resemblance to either the climatological circulation or AM4’s response. This pronounced discrepancy underscores a fundamental limitation: these ML models not only misrepresent the mean meridional mass transport in the present climate but also exhibit dynamically implausible responses under warming (e.g., latitudes with net mass flux through the surface). NeuralGCM's mean overturning circulation and its response to uniform SST warming also deviate substantially from AM4. Overall, this basic feature of the atmospheric circulation is one that warrants further attention in ML models to improve their ability to simulate realistic overturning circulations.

The radiative responses further underscore differences among the models. While ACE2 and AM4 broadly agree on the sign of radiative changes, ACE2 exhibits weaker increases in OLR than AM4. cBottle even produces an unstable net radiative feedback, largely due to reduced upward SW radiation (i.e., an increase in absorbed solar radiation that exceeds the increase in OLR). The extent to which these radiative responses are out of the range of GCM simulated feedbacks depends in part on whether one considers the input SST as the relevant temperature change or the output surface air temperature: ACE2 and cBottle have weak increases in OLR and weak surface air warming. Further research on interpreting the controlling factors for the ML model radiative fluxes and how they accord with physical understanding would be instructive (cf. \cite{hafner25}). 

Based on the analyses presented in this study, it is evident that NeuralGCM’s hybrid design, which combines a dynamical core with neural-network-based parameterizations, more closely matches well understood physical responses to warming. Its structure also enables it to reproduce dynamically consistent responses, such as amplified upper-tropospheric warming and strengthened westerlies. In contrast, the cBottle and ACE2 design of ``whole-model'' emulators that infer the full atmospheric state directly from boundary forcing do not extrapolate to perturbed climates when trained on reanalysis states. 
The extent to which physical constraints beyond the global-mean conservation laws used in ACE2's architecture can improve generalization is an important question for advancing the physical realism and robustness of next-generation ML climate models.

One motivation for the model comparison performed here is to establish a baseline of how well current model architectures and training strategies generalize to the out-of-sample climate changes of interest. To that end, it is valuable to examine both the overall response in quantities of interest and the inter-variable dependencies, which motivated our analysis of variables like column water vapor and the meridional circulation that are known to affect precipitation. Inter-variable relationships have also recently been used as part of the architecture for ML weather prediction \cite{fan25}. 
It is a challenge to solely use the reanalysis-era for training to build a model that captures the response to SST warming, though there are efforts to design climate-invariant ML approaches that have been successful for parameterizations in idealized simulations \cite{beucler2024climate,liu25} and to develop coarse-grained, physically informed models \cite{falasca2025neural}. \add{As discussed in the introduction, our analysis focuses on metrics that exhibit robust behavior across physical models, supported by relatively high confidence and clear physical explanations. We emphasize, however, that the response of the physical model used in this study should not be regarded as ground truth.}


Our results highlight the promise of ML models as complementary tools to physical models for climate research, such as the precipitation response to warming.  
Identifying the tradeoffs between the low computational cost and fidelity of the response is an important validation step for current ML global atmospheric models. 
Uniform SST warming experiments provide a tractable and informative benchmark for ML models simulating climate change. There is the potential for significant progress from today's frontier to improve their robustness and reliability in applications involving future climate change scenarios.

\section*{Open Research Section}
The cBottle code is available at https://github.com/NVlabs/cBottle, ACE2-ERA5 at https://github.com/ai2cm/ace, NeuralGCM at https://neuralgcm.readthedocs.io/en/latest/index.html, and AM4 at https://github.com/NOAA-GFDL/AM4. The code for generating figures for this study can be accessed via this link \url{https://tigress-web.princeton.edu/~bosongz/emulators/code/}. The simulations were conducted and archived on Princeton’s computing system. The datasets generated and analyzed during the current study are available in the \url{https://tigress-web.princeton.edu/~bosongz/emulators/}. 

\section*{Conflict of Interest declaration}
The authors declare no competing financial or non-financial interests.

\acknowledgments

We thank Spencer Clark, Will Gregory, and Paul O'Gorman for helpful discussions. We thank Janni Yuval for disclosing NeuralGCM's hardware-dependent behavior and constructive suggestions for how to address this.
Simulations presented here were performed using High Performance Computing resources provided by the Cooperative Institute for Modeling the Earth System, with help from the Princeton Institute for Computational Science and Engineering. We acknowledge support from the National Oceanic and Atmospheric Administration, U.S. Department of Commerce under award NA23OAR4320198. The statements, findings, conclusions, and recommendations are those of the author and do not necessarily reflect the views of the National Oceanic and Atmospheric Administration, or the U.S. Department of Commerce.


%
%

\bibliography{lit_tm}

@misc{NASA/LARC/SD/ASDC2024CERES,
	publisher={NASA Langley Atmospheric Science Data Center DAAC},
	title={CERES Energy Balanced and Filled (EBAF) TOA Monthly means data in netCDF Edition4.2.1},
	url={https://doi.org/10.5067/TERRA-AQUA-NOAA20/CERES/EBAF-TOA_L3B004.2.1},
	author={NASA/LARC/SD/ASDC},
	date={2024-12-02},
	year=2024,
	month=12,
	day=2,
}

@article{rackow24,
  title={Robustness of {AI}-based weather forecasts in a changing climate},
  author={Rackow, Thomas and Koldunov, Nikolay and Lessig, Christian and Sandu, Irina and Alexe, Mihai and Chantry, Matthew and Clare, Mariana and Dramsch, Jesper and Pappenberger, Florian and Pedruzo-Bagazgoitia, Xabier and others},
  journal={arXiv preprint arXiv:2409.18529},
  year={2024}
}

@article{landsberg25,
  title={Forecasting the Future with Yesterday's Climate: {T}emperature Bias in {AI} Weather and Climate Models},
  author={Landsberg, J. B. and Barnes, E. A.},
  journal={arXiv preprint arXiv:2509.22359},
  year={2025}
}

@article{fan25,
  title={Incorporating Multivariate Consistency in {M}-Based Weather Forecasting with Latent-space Constraints},
  author={Fan, Hang and Xiao, Yi and Qu, Yongquan and Ling, Fenghua and Fei, Ben and Bai, Lei and Gentine, Pierre},
  journal={arXiv preprint arXiv:2510.04006},
  year={2025}
}

@article{hafner25,
  title={Interpretable machine learning-based radiation emulation for {ICON}},
  author={Hafner, Katharina and Iglesias-Suarez, Fernando and Shamekh, Sara and Gentine, Pierre and Giorgetta, Marco A and Pincus, Robert and Eyring, Veronika},
  journal={Journal of Geophysical Research: Machine Learning and Computation},
  volume={2},
  number={4},
  pages={e2024JH000501},
  year={2025},
  publisher={Wiley Online Library}
}

@article{Allen02,
        Author = {M. R. Allen and W. J. Ingram},
        Journal = {Nature},
        Pages = {224--232},
        Title = {Constraints on future changes in climate and the hydrologic cycle},
        Volume = 419,
        Year = 2002
}

@article{baxter25,
  title={Benchmarking atmospheric circulation variability in an {AI} emulator, {ACE2}, and a hybrid model, {NeuralGCM}},
  author={Baxter, Ian and Pahlavan, Hamid and Hassanzadeh, Pedram and Rucker, Katharine and Shaw, Tiffany},
  journal={arXiv preprint arXiv:2510.04466},
  year={2025}
}

@article{zhang2023using,
  title={Using a Green’s function approach to diagnose the pattern effect in GFDL AM4 and CM4},
  author={Zhang, Bosong and Zhao, Ming and Tan, Zhihong},
  journal={Journal of Climate},
  volume={36},
  number={4},
  pages={1105--1124},
  year={2023}
}

@article{bloch2024green,
  title={The green's function model intercomparison project (GFMIP) protocol},
  author={Bloch-Johnson, Jonah and Rugenstein, Maria AA and Alessi, Marc J and Proistosescu, Cristian and Zhao, Ming and Zhang, Bosong and Williams, Andrew IL and Gregory, Jonathan M and Cole, Jason and Dong, Yue and others},
  journal={Journal of Advances in Modeling Earth Systems},
  volume={16},
  number={2},
  pages={e2023MS003700},
  year={2024},
  publisher={Wiley Online Library}
}

@article{falasca2025neural,
  title={Neural models of multiscale systems: conceptual limitations, stochastic parametrizations, and a climate application},
  author={Falasca, Fabrizio},
  journal={arXiv preprint arXiv:2506.22552},
  year={2025}
}

@article{lionello2024hadley,
  title={The Hadley circulation in a changing climate},
  author={Lionello, Piero and D'Agostino, Roberta and Ferreira, David and Nguyen, Hanh and Singh, Martin S},
  journal={Annals of the New York Academy of Sciences},
  volume={1534},
  number={1},
  pages={69--93},
  year={2024},
  publisher={Wiley Online Library}
}

@article{vecchi2007global,
  title={Global warming and the weakening of the tropical circulation},
  author={Vecchi, Gabriel A and Soden, Brian J},
  journal={Journal of Climate},
  volume={20},
  number={17},
  pages={4316--4340},
  year={2007}
}

@article{d2017factors,
  title={Factors controlling Hadley circulation changes from the Last Glacial Maximum to the end of the 21st century},
  author={D'Agostino, Roberta and Lionello, Piero and Adam, Ori and Schneider, Tapio},
  journal={Geophysical Research Letters},
  volume={44},
  number={16},
  pages={8585--8591},
  year={2017},
  publisher={Wiley Online Library}
}

@article{liu25,
  title={{CERA}: A Framework for Improved Generalization of Machine Learning Models to Changed Climates},
  author={Liu, S. and O'Gorman, P. A.},
  journal={arXiv preprint arXiv:2509.00010},
  year={2025}
}

@article{kushner01,
  title={Southern Hemisphere atmospheric circulation response to global warming},
  author={Kushner, P. J. and Held, I. M. and Delworth, T. L.},
  journal={J. Climate},
  volume={14},
  pages={2238--2249},
  year={2001}
}

@article{bordoni2025futures,
  title={The futures of climate modeling},
  author={Bordoni, S and Kang, SM and Shaw, Tiffany A and Simpson, IR and Zanna, L},
  journal={npj Climate and Atmospheric Science},
  volume={8},
  number={1},
  pages={99},
  year={2025},
  publisher={Nature Publishing Group UK London}
}

@article{dheeshjith2025samudra,
  title={Samudra: An AI global ocean emulator for climate},
  author={Dheeshjith, Surya and Subel, Adam and Adcroft, Alistair and Busecke, Julius and Fernandez-Granda, Carlos and Gupta, Shubham and Zanna, Laure},
  journal={Geophysical Research Letters},
  volume={52},
  number={10},
  pages={e2024GL114318},
  year={2025},
  publisher={Wiley Online Library}
}

@article{clark2024ace2,
  title={{ACE2-SOM}: Coupling an ML atmospheric emulator to a slab ocean and learning the sensitivity of climate to changed CO$_2$},
  author={Clark, Spencer K and Watt-Meyer, Oliver and Kwa, Anna and McGibbon, Jeremy and Henn, Brian and Perkins, W Andre and Wu, Elynn and Harris, Lucas M and Bretherton, Christopher S},
  journal={arXiv preprint arXiv:2412.04418},
  year={2024}
}

@article{chapman2025camulator,
  title={CAMulator: Fast emulation of the community atmosphere model},
  author={Chapman, William E and Schreck, John S and Sha, Yingkai and Gagne II, David John and Kimpara, Dhamma and Zanna, Laure and Mayer, Kirsten J and Berner, Judith},
  journal={arXiv preprint arXiv:2504.06007},
  year={2025}
}

@article{neelin2022precipitation,
  title={Precipitation extremes and water vapor: Relationships in current climate and implications for climate change},
  author={Neelin, J David and Martinez-Villalobos, Cristian and Stechmann, Samuel N and Ahmed, Fiaz and Chen, Gang and Norris, Jesse M and Kuo, Yi-Hung and Lenderink, Geert},
  journal={Current Climate Change Reports},
  volume={8},
  number={1},
  pages={17--33},
  year={2022},
  publisher={Springer}
}

@article{kuo2020convective,
  title={Convective transition statistics over tropical oceans for climate model diagnostics: GCM evaluation},
  author={Kuo, Yi-Hung and Neelin, J David and Chen, Chih-Chieh and Chen, Wei-Ting and Donner, Leo J and Gettelman, Andrew and Jiang, Xianan and Kuo, Kuan-Ting and Maloney, Eric and Mechoso, Carlos R and others},
  journal={Journal of the Atmospheric Sciences},
  volume={77},
  number={1},
  pages={379--403},
  year={2020}
}

@article{beucler2024climate,
  title={Climate-invariant machine learning},
  author={Beucler, Tom and Gentine, Pierre and Yuval, Janni and Gupta, Ankitesh and Peng, Liran and Lin, Jerry and Yu, Sungduk and Rasp, Stephan and Ahmed, Fiaz and O’Gorman, Paul A and others},
  journal={Science Advances},
  volume={10},
  number={6},
  pages={eadj7250},
  year={2024},
  publisher={American Association for the Advancement of Science}
}

@article{lam2023learning,
  title={Learning skillful medium-range global weather forecasting},
  author={Lam, Remi and Sanchez-Gonzalez, Alvaro and Willson, Matthew and Wirnsberger, Peter and Fortunato, Meire and Alet, Ferran and Ravuri, Suman and Ewalds, Timo and Eaton-Rosen, Zach and Hu, Weihua and others},
  journal={Science},
  volume={382},
  number={6677},
  pages={1416--1421},
  year={2023},
  publisher={American Association for the Advancement of Science}
}

@article{bi2023accurate,
  title={Accurate medium-range global weather forecasting with 3D neural networks},
  author={Bi, Kaifeng and Xie, Lingxi and Zhang, Hengheng and Chen, Xin and Gu, Xiaotao and Tian, Qi},
  journal={Nature},
  volume={619},
  number={7970},
  pages={533--538},
  year={2023},
  publisher={Nature Publishing Group UK London}
}

@article{rasp2024weatherbench,
  title={WeatherBench 2: A benchmark for the next generation of data-driven global weather models},
  author={Rasp, Stephan and Hoyer, Stephan and Merose, Alexander and Langmore, Ian and Battaglia, Peter and Russell, Tyler and Sanchez-Gonzalez, Alvaro and Yang, Vivian and Carver, Rob and Agrawal, Shreya and others},
  journal={Journal of Advances in Modeling Earth Systems},
  volume={16},
  number={6},
  pages={e2023MS004019},
  year={2024},
  publisher={Wiley Online Library}
}

@article{keisler2022forecasting,
  title={Forecasting global weather with graph neural networks},
  author={Keisler, Ryan},
  journal={arXiv preprint arXiv:2202.07575},
  year={2022}
}

@article{bracco2025machine,
  title={Machine learning for the physics of climate},
  author={Bracco, Annalisa and Brajard, Julien and Dijkstra, Henk A and Hassanzadeh, Pedram and Lessig, Christian and Monteleoni, Claire},
  journal={Nature Reviews Physics},
  volume={7},
  number={1},
  pages={6--20},
  year={2025},
  publisher={Nature Publishing Group UK London}
}

@article{eyring2024ai,
  title={AI-empowered next-generation multiscale climate modelling for mitigation and adaptation},
  author={Eyring, Veronika and Gentine, Pierre and Camps-Valls, Gustau and Lawrence, David M and Reichstein, Markus},
  journal={Nature Geoscience},
  volume={17},
  number={10},
  pages={963--971},
  year={2024},
  publisher={Nature Publishing Group UK London}
}

@article{huffman2023new,
  title={The new version 3.2 Global Precipitation Climatology Project (GPCP) monthly and daily precipitation products},
  author={Huffman, George J and Adler, Robert F and Behrangi, Ali and Bolvin, David T and Nelkin, Eric J and Gu, Guojun and Ehsani, Mohammad Reza},
  journal={Journal of Climate},
  volume={36},
  number={21},
  pages={7635--7655},
  year={2023},
  publisher={American Meteorological Society}
}

@article{brenowitz2025climate,
  title={Climate in a bottle: Towards a generative foundation model for the kilometer-scale global atmosphere},
  author={Brenowitz, Noah D and Ge, Tao and Subramaniam, Akshay and Gupta, Aayush and Hall, David M and Mardani, Morteza and Vahdat, Arash and Kashinath, Karthik and Pritchard, Michael S},
  journal={arXiv preprint arXiv:2505.06474},
  year={2025}
}

@article{kochkov2024neural,
  title={Neural general circulation models for weather and climate},
  author={Kochkov, Dmitrii and Yuval, Janni and Langmore, Ian and Norgaard, Peter and Smith, Jamie and Mooers, Griffin and Kl{\"o}wer, Milan and Lottes, James and Rasp, Stephan and D{\"u}ben, Peter and others},
  journal={Nature},
  volume={632},
  number={8027},
  pages={1060--1066},
  year={2024},
  publisher={Nature Publishing Group UK London}
}

@article{watt2025ace2,
  title={ACE2: accurately learning subseasonal to decadal atmospheric variability and forced responses},
  author={Watt-Meyer, Oliver and Henn, Brian and McGibbon, Jeremy and Clark, Spencer K and Kwa, Anna and Perkins, W Andre and Wu, Elynn and Harris, Lucas and Bretherton, Christopher S},
  journal={npj Climate and Atmospheric Science},
  volume={8},
  number={1},
  pages={205},
  year={2025},
  publisher={Nature Publishing Group UK London}
}

@article{watt2023ace,
  title={ACE: A fast, skillful learned global atmospheric model for climate prediction},
  author={Watt-Meyer, Oliver and Dresdner, Gideon and McGibbon, Jeremy and Clark, Spencer K and Henn, Brian and Duncan, James and Brenowitz, Noah D and Kashinath, Karthik and Pritchard, Michael S and Bonev, Boris and others},
  journal={arXiv preprint arXiv:2310.02074},
  year={2023}
}

@article{wu2025applying,
  title={Applying the ACE2 Emulator to SST Green's Functions for the E3SMv3 Global Atmosphere Model},
  author={Wu, Elynn and Rebassoo, Finn and Paul, Pappu and Proistosescu, Cristian and Nugent, Jacqueline and McCoy, Daniel and Caldwell, Peter and Bretherton, Christopher S},
  journal={arXiv preprint arXiv:2505.08742},
  year={2025}
}

@article{zhao2018gfdlam4part1,
  title={The GFDL global atmosphere and land model AM4. 0/LM4. 0: 1. Simulation characteristics with prescribed SSTs},
  author={Zhao, Ming and Golaz, J-C and Held, IM and Guo, H and Balaji, V and Benson, R and Chen, J-H and Chen, X and Donner, LJ and Dunne, JP and others},
  journal={Journal of Advances in Modeling Earth Systems},
  volume={10},
  number={3},
  pages={691--734},
  year={2018},
  publisher={Wiley Online Library}
}

@article{zhao2018gfdlam4part2,
  title={The GFDL global atmosphere and land model AM4. 0/LM4. 0: 2. Model description, sensitivity studies, and tuning strategies},
  author={Zhao, Ming and Golaz, J-C and Held, IM and Guo, H and Balaji, V and Benson, R and Chen, J-H and Chen, X and Donner, LJ and Dunne, JP and others},
  journal={Journal of Advances in Modeling Earth Systems},
  volume={10},
  number={3},
  pages={735--769},
  year={2018},
  publisher={Wiley Online Library}
}

@article{yuval2024neural,
  title={Neural general circulation models optimized to predict satellite-based precipitation observations},
  author={Yuval, Janni and Langmore, Ian and Kochkov, Dmitrii and Hoyer, Stephan},
  journal={arXiv preprint arXiv:2412.11973},
  year={2024}
}

@article{yuval2026neural,
  title={Neural general circulation models for modeling precipitation},
  author={Yuval, Janni and Langmore, Ian and Kochkov, Dmitrii and Hoyer, Stephan},
  journal={Science Advances},
  volume={12},
  number={2},
  pages={eadv6891},
  year={2026},
  publisher={American Association for the Advancement of Science}
}

@article{henn2026aimip,
  title={AIMIP Phase 1: systematic evaluations of AI weather and climate models},
  author={Henn, Brian and Bretherton, Christopher S and Kodunov, Nikolay and Lessig, Christian and Molina, Maria J and Arcomano, Troy and Watt-Meyer, Oliver and Couairon, Guillaume and Singh, Renu and Brunstein, Robert and others},
  journal={arXiv preprint arXiv:2605.06944},
  year={2026}
}

@article{rucker2026benchmarking,
  title={Benchmarking Regional Thermodynamic Trends in an AI emulator, ACE2, and a hybrid model, NeuralGCM},
  author={Rucker, Katharine and Baxter, Ian and Hassanzadeh, Pedram and Shaw, Tiffany A and Pahlavan, Hamid A},
  journal={Geophysical Research Letters},
  volume={53},
  number={10},
  pages={e2025GL120185},
  year={2026},
  publisher={Wiley Online Library}
}

@article{chen2026hierarchical,
  title={Hierarchical testing of a hybrid machine learning-physics global atmosphere model},
  author={Chen, Ziming and Leung, L Ruby and Zhou, Wenyu and Lu, Jian and Lubis, Sandro W and Liu, Ye and Chang, Chuan-Chieh and Harrop, Bryce E and Wang, Ya and Yang, Mingshi and others},
  journal={AGU Advances},
  volume={7},
  number={2},
  pages={e2025AV002075},
  year={2026},
  publisher={Wiley Online Library}
}

@article{bretherton04b,
  title={Relationships between water vapor path and precipitation over the tropical oceans},
  author={Bretherton, C. S. and Peters, M. E. and Back, L. E.},
  journal={J. Climate},
  volume={17},
  pages={1517--1528},
  year={2004}
}

@article{bretherton22,
  title={Correcting Coarse-Grid Weather and Climate Models by Machine Learning From Global Storm-Resolving Simulations},
  author={Bretherton, C. S. and Henn, B. and Kwa, A. and Brenowitz, N. D. and Watt-Meyer, O. and McGibbon, J. and Perkins, W. A. and Clark, S. K. and Harris, L.},
  journal={J. Adv. Model. Earth Syst.},
  volume={14},
  number={2},
  pages={e2021MS002794},
  year={2022},
  publisher={Wiley Online Library}
}

@article{byrne13a,
  title={Land-Ocean Warming Contrast over a Wide Range of Climates: Convective Quasi-Equilibrium Theory and Idealized Simulations},
  author={Byrne, M. P. and O'Gorman, P. A.},
  journal={J. Climate},
  volume={26},
  pages={4000--4016},
  year={2013},
}

@article{byrne13b,
  title={Link between land-ocean warming contrast and surface relative humidities in simulations with coupled climate models},
  author={Byrne, M. P. and O'Gorman, P. A.},
  journal={Geophys. Res. Lett.},
  year={2013},
  publisher={Wiley Online Library}
}

@article{cess90,
  title={Intercomparison and interpretation of climate feedback processes in 19 atmospheric general circulation models},
  author={Cess, R. D. and Potter, G. L. and Blanchet, J.-P. and Boer, G. J. and Del Genio, A. D. and Deque, M. and Dymnikov, V. and Galin, V. and Gates, W. L. and Ghan, S. J. and others},
  journal={J. Geophys. Res.},
  volume={95},
  pages={16601--16615},
  year={1990},
  publisher={Wiley Online Library}
}

@article{eyring16,
  title={Overview of the Coupled Model Intercomparison Project Phase 6 ({CMIP6}) experimental design and organization},
  author={Eyring, V. and Bony, S. and Meehl, G. A. and Senior, C. A. and Stevens, B. and Stouffer, R. J. and Taylor, K. E.},
  journal={Geosci. Model Dev.},
  volume={9},
  pages={1937--1958},
  year={2016},
  publisher={Copernicus GmbH}
}

@article{Hartmann02,
  title={An important constraint on tropical cloud-climate feedback},
  author={Hartmann, D. L. and Larson, K.},
  journal={Geophys. Res. Lett.},
  volume={29},
  pages = {1951},
  year={2002},
  note={doi:10.1029/2002GL015835},
  publisher={American Geophysical Union}
}

@article{held06,
	Author = {I. M. Held and B. J. Soden},
	Journal = {J. Climate},
	Pages = {5686--5699},
	Title = {Robust responses of the hydrological cycle to global warming},
	Volume = 19,
	Year = 2006
}

@article{hersbach20,
  title={The {ERA5} global reanalysis},
  author={Hersbach, H. and Bell, B. and Berrisford, Paul and Hirahara, Shoji and Hor{\'a}nyi, Andr{\'a}s and Mu{\~n}oz-Sabater, Joaqu{\'\i}n and Nicolas, Julien and Peubey, Carole and Radu, Raluca and Schepers, Dinand and others},
  journal={Quart. J. Roy. Meteor. Soc.},
  volume={146},
  pages={1999--2049},
  year={2020},
  publisher={Wiley Online Library}
}

@article{jeevanjee18,
  title={Mean precipitation change from a deepening troposphere},
  author={Jeevanjee, N. and Romps, D. M.},
  journal={Proc. Nat. Acad. Sci.},
  volume={115},
  pages={11465--11470},
  year={2018},
  publisher={National Acad Sciences}
}

@article{joshi08,
  title={Mechanisms for the land/sea warming contrast exhibited by simulations of climate change},
  author={Joshi, M. M. and Gregory, J. M. and Webb, M. J. and Sexton, D. M. H. and Johns, T. C.},
  journal={Clim. Dyn.},
  volume={30},
  pages={455--465},
  year={2008},
  publisher={Springer}
}

@article{merlis24,
  title={Climate Sensitivity and Relative Humidity Changes in Global Storm-Resolving Model Simulations of Climate Change},
  author={Merlis, T. M. and Cheng, K.-Y. and Guendelman, I. and Harris, L. and Bretherton, C. S. and Bolot, M. and Zhou, L. and Kaltenbaugh, A. and Clark, S. and Vecchi, G. A. and Fueglistaler, S.},
  journal={Sci. Adv.},
  volume={10},
  pages={eadn5217},
  year={2024}
}

@article{OGorman09a,
	Author = {P. A. O'Gorman and T. Schneider},
	Journal = {J. Climate},
	Pages = {5676--5685},
	Title = {Scaling of precipitation extremes over a wide range of climates simulated with an idealized {GCM}},
	Volume = {22},
	Year = 2009
}

@article{OGorman09b,
	Author = {P. A. O'Gorman and T. Schneider},
	Journal = {Proc. Nat. Acad. Sci.},
	Pages = {14773--14777},
	Title = {The physical basis for increases in precipitation extremes in simulations of 21st-century climate change},
	Volume = {106},
	Year = 2009
}

@article{santer05,
        Author = {Santer, B. D. and Wigley, T. M. L. and Mears, C. and Wentz, F. J. and Klein, S. A. and Seidel, D. J. and Taylor, K. E. and Thorne, P. W. and Wehner, M. F. and Gleckler, P. J. and others},
        Journal = {Science},
        Pages = {1551--1556},
        Title = {Amplification of surface temperature trends and variability in the tropical atmosphere},
        Volume = {309},
        Year = {2005}
}

@article{webb17,
  title={The cloud feedback model intercomparison project ({CFMIP}) contribution to {CMIP6}},
  author={Webb, M. J. and Andrews, T. and Bodas-Salcedo, A. and Bony, S. and Bretherton, C. S. and Chadwick, R. and Chepfer, H. and Douville, H. and Good, P. and Kay, J. E. and others},
  journal={Geosci. Model Dev.},
  volume={10},
  pages={359--384},
  year={2017},
  publisher={Copernicus GmbH}
}

%
%
%
%
%

\end{document}